# Effective and Complete Discovery of Order Dependencies via Set-based Axiomatization

Jaroslaw Szlichta[#], Parke Godfrey[*], Lukasz Golab[$], Mehdi Kargar[♮], Divesh Srivastava[◇]
[#]University of Ontario Institute of Technology, Canada
[*]York University, Canada
[$]University of Waterloo, Canada
[♮]University of Windsor, Canada
[◇]AT&T Labs-Research, USA
jaroslaw.szlichta@uoit.ca, godfrey@yorku.ca, lgolab@uwaterloo.ca, mkargar@uwindsor.ca, divesh@research.att.com

## ABSTRACT

Integrity constraints (ICs) provide a valuable tool for expressing and enforcing application semantics. However, formulating constraints manually requires domain expertise, is prone to human errors, and may be excessively time consuming, especially on large datasets. Hence, proposals for automatic discovery have been made for some classes of ICs, such as functional dependencies (FDs), and recently, order dependencies (ODs). ODs properly subsume FDs, as they can additionally express business rules involving order; e.g., an employee never has a higher salary while paying lower taxes compared with another employee.

We address the limitations of prior work on OD discovery which has factorial complexity in the number of attributes, is incomplete (i.e., it does not discover valid ODs that cannot be inferred from the ones found) and is not concise (i.e., it can result in "redundant" discovery and overly large discovery sets). We improve significantly on complexity, offer completeness, and define a compact canonical form. This is based on a novel polynomial mapping to a canonical form for ODs, and a sound and complete set of axioms (inference rules) for canonical ODs. This allows us to develop an efficient set-containment, lattice-driven OD discovery algorithm that uses the inference rules to prune the search space. Our algorithm has exponential worst-case time complexity, $O(2^{|\mathbf{R}|})$, in the number of attributes and linear complexity in the number of tuples. We prove that it produces a complete, minimal set of ODs (i.e., minimal with regards to the canonical representation). Finally, using real and synthetic datasets, we experimentally show orders-of-magnitude performance improvements over the current state-of-the-art algorithm and demonstrate effectiveness of our techniques.

## 1. INTRODUCTION

### 1.1 Motivation

Table 1: A table with employee salaries and tax information.

| # | ID | yr | posit | bin | sal | perc | tax | grp | subg |
|---|---|---|---|---|---|---|---|---|---|
| *t1* | 10 | 16 | secr | 1 | 5K | 20% | 1K | A | III |
| *t2* | 11 | 16 | mngr | 2 | 8K | 25% | 2K | C | II |
| *t3* | 12 | 16 | direct | 3 | 10K | 30% | 3K | D | I |
| *t4* | 10 | 15 | secr | 1 | 4.5K | 20% | 0.9K | A | III |
| *t5* | 11 | 15 | mngr | 2 | 6K | 25% | 1.5K | C | I |
| *t6* | 12 | 15 | direct | 3 | 8K | 25% | 2K | C | II |

With the interest in data analytics at an all-time high, *data quality* and *query optimization* are being revisited to address the scale and complexity of modern data-intensive applications. Real data suffer from inconsistencies, duplicates and missing values [14, 15, 19]. A recent Gartner Research Report study [12] revealed that, by 2017, one third of the largest global companies will experience data quality crises due to their inability to trust and govern their enterprise information. Deep analytics on large data warehouses, spanning thousands of lines of SQL code, are no longer restricted to well-tuned, canned batch reports [5]. Instead, complex ad-hoc queries are increasingly required for business operations to make timely data-driven decisions. Without clean data and effective query optimization, organizations will not be able to stay competitive and to take advantage of new data-driven opportunities.

Integrity constraints (ICs), which specify the intended semantics and attribute relationships, are commonly used to characterize data quality and to optimize business queries. Prior work has focused on *functional dependencies* (FDs) and extensions thereof, such as *conditional* FDs [14]. FDs cannot capture relationships among *ordered* attributes, however, such as between timestamps and numbers, which are common in business data [24]. For example, consider Table 1, which shows employee tax records in which tax is calculated as a percentage (perc) of salary (sal). Both tax and percentage increase with salary.

We study order dependencies (ODs) [22, 25], which can naturally express such semantics. The OD salary *orders* tax, percentage holds: if we sort the table by salary, we know it is also sorted by tax, percentage; i.e., sorted by tax, with ties broken by percentage. Similarly, the OD salary *orders* group (grp), subg holds in Table 1. The class of ODs subsumes the class of FDs [22]. For example, if salary *orders* tax, percentage, then the FD that salary *functionally determines* tax, percentage must also hold.

```
select D.d_year, D.d_quarter,
    D.d_month, sum(WS.sales) as total,
    count(*) as quantity
from web_sales WS, date_dim D,
where WS.date_sk = D.date_sk and
    total > 1,000,000 and
    D.d_year between
        2012 and 2016
group by D.d_year, D.d_quarter,
    D.d_month
order by D.d_year, D.d_quarter,
    D.d_month;
```

Query 1. Example of query optimization by ODs.

The additional expressiveness of ODs makes them particularly suitable for improving data quality, where ODs can describe intended semantics and business rules; and their violations point out possible data errors. ODs enable new optimization techniques for *online analytical processing* (OLAP) queries: eliminating expensive joins from query plans in data warehouse environments [23]; processing queries with order-by, group-by, distinct, and partition statements [24, 25]; and improving the performance of queries with SQL functions and algebraic expressions (e.g., `date` *orders* `year(date)` and `date` *orders* `year(date)*100 + month(date)`) [16, 25].

We illustrate some possible optimizations using Query 1 over a data warehouse schema such as the TPC-DS[1] benchmark. Financial year is represented by d_year, d_quarter and d_month, and is stored in the date_dim table. The values of quarter divide the financial year into four three-month periods. The fact table web_sales has a foreign key d_date_sk that references the dimension table date_dim, which is a surrogate key, a sequential number, assigned with increasing values as data is added. Assume date_dim has an index on d_year and d_month. Since d_month *functionally determines* d_quarter, the group-by on d_year, d_quarter, d_month is semantically equivalent to that on d_year, d_month. In fact, some optimizers would eliminate fiscal quarter via the relevant FDs [21]. However, this FD cannot simplify the order-by statement to match the given index. For this, we would need an OD d_month *orders* d_quarter [24, 25].

In a data warehouse, most queries reference fact tables which include foreign keys to dimension tables. These foreign keys are often surrogate keys. However, queries usually reference other dimension attributes, not surrogate keys; e.g., the "between" predicate on d_year in Query 1. This predicate may require a potentially expensive join between the fact table and the date dimension table to identify all surrogate key values falling between year 2012 and year 2016. However, if we know that d_date_sk *orders* d_year, then it suffices to make two probes into the date dimension table: one to find the minimum relevant d_date_sk value corresponding to January 1 2012; and one to find the maximum relevant d_date_sk value corresponding to December 31 2016. This allows us to restate the "between" predicate in terms of these two surrogate key values, rather than by years, therefore eliminating the join [23].

With the help of ODs, query optimizers can eliminate costly operators such as joins and sorts, and can identify additional *interesting orders*: ordered streams between query operators that exploit the available indices, enable pipelining, and eliminate intermediate sorts and partitioning steps [25]. Sorting and interesting orders are integral parts of relational query optimizers, not only for SQL order-by and group-by, but for instance also for sort-merge joins [24] and implementing indexes [6].

Date and time are richly supported in the SQL standard and frequently appear in decision support queries over historical data. For example, the widely-used TPC-DS benchmark contains 99 queries, of which 85 involve date operators and predicates and five involve time operators and predicates. If the concept of ODs were only applicable to date and time, they would already be very beneficial. As seen in Table 1, ODs are also useful in other domains arising from business semantics over many types of measures such as sales, flight schedules, stock prices, salaries and taxes [22, 24, 25].

## 1.2 Problem Statement and Contributions

To use ODs for data cleaning and query optimization as outlined above, we need to know which ODs hold on a given dataset. While ODs can be obtained manually through consultation with domain experts, this is known to be an expensive, time consuming, and error-prone process that requires expertise in the data dependency language [11]. The problem we study in this paper is *how to discover automatically ODs from data.* Such automatically discovered ODs can be manually validated by domain experts, which is an easier task than manual specification.

This problem has been studied before, but is not well understood. Our aim is to provide a deeper understanding of OD discovery. An OD discovery algorithm was recently proposed by Langer and Naumann [13], which has a *factorial* worst-case time complexity in the number of attributes. (This is the only prior OD discovery work of which we are aware.) In contrast to FDs, ODs are naturally expressed with lists rather than sets of attributes. The first sound and complete axiomatization for ODs is expressed in a list notation [22]; Langer and Naumann use this. For instance, salary *orders* tax, percentage is different from salary *orders* percentage, tax whereas the two analogous FDs are equivalent.

An insight we present is that ODs can be expressed with sets of attributes via a *polynomial mapping* into a set-based canonical form. The mapping allows us to design a *fast* and *effective* OD discovery algorithm that has *exponential* worst-case complexity, $O(2^{|\mathbf{R}|})$, in the number of attributes $|\mathbf{R}|$ (and linear complexity in the number of tuples). This complexity is similar to previous FD and inclusion dependency discovery algorithms such as TANE [10]. In one of our experimental runs, using a flight dataset with 1000 tuples and 20 attributes, our algorithm takes under one second whereas the algorithm from [13] does not terminate after five hours (Section 5).

We also develop *sound* and *complete* set-based axioms (inference rules) for ODs that enable *pruning* of the search space, which can alleviate the worst-case complexity in practice. Our pruning rules do not affect the completeness of the discovered ODs. The list-based algorithm in [13] uses aggressive pruning rules to overcome their factorial complexity, which we show leads to incompleteness (counter to the claim of complete discovery made in [13]). They can miss important ODs. For example, the sub-class of ODs in which the same attribute(s) are repeated in the left-hand-side and the right-hand-side such as year(yr), salary *orders* year, bin is missed (see Section 4.5). Their canonical form designed to accommodate a list-based lattice, is incomplete for representing ODs.

[1] http://tpc.org/tpcds/

Lastly, the list-based canonical form for ODs used in [13] is not compact. By introducing a set-based canonical form for ODs, we can achieve much greater compactness in our representation. Many ODs that are considered minimal by the algorithm in [13] are found to be redundant by our algorithm. This is quite important for efficiency in OD discovery. We do not need to rediscover the "same" ODs repeatedly–that is, ODs that can be inferred from ones we have already discovered–and we can more aggressively prune portions of the search space which would only have "repeats".

In Section 2, we define ODs and provide additional examples. In this paper, we make the following contributions.

1. *Mapping.* We translate ODs (formally defined in Section 2) into a novel set-based *canonical form* (Section 3) which leads to a new and efficient approach to OD discovery (Section 4). By mapping ODs into *equivalent* set-based *canonical* ODs, we illustrate that they can be discovered efficiently by traversing a *set-containment lattice* with *exponential* worst-case complexity in the number of attributes (the same as for FDs [11]), and just *linear* in the number of tuples (Section 4.7).
2. *Set-based Axiomatization.* We introduce axioms for set-based canonical ODs and prove these are *sound* and *complete* (Section 3.2). Our inference rules reveal insights into the nature of canonical ODs, which lead to optimizations of the OD discovery algorithm to avoid redundant computation (Section 4).
3. *Completeness and Minimality.* We prove that our discovery algorithm produces a *complete*, *minimal* set of ODs (Section 4).
4. *Experiments.* We provide an experimental study (Section 5) of the performance and effectiveness of our discovery techniques using real datasets. We demonstrate the speedup due to our pruning rules, and report orders-of-magnitude performance improvements over previous work [13].

We discuss related work in Section 6. In Section 7, we conclude and consider future work.

# 2. PRELIMINARIES

We begin by formally defining ODs, present the established list-based axiomatization [22], and explain the two ways in which they can be violated.

## 2.1 Framework and Background

We use the following notational conventions.

- **Relations.** $\mathbf{R}$ denotes a *relation* (*schema*) and $\mathbf{r}$ denotes a specific *relation instance* (*table*). A, B and C denote single *attributes*, $s$ and $t$ denote *tuples*, and $t_\mathsf{A}$ denotes the value of an attribute A in a tuple $t$.
- **Sets.** $\mathcal{X}$ and $\mathcal{Y}$ denote *sets* of attributes. Let $t_\mathcal{X}$ denote the *projection* of tuple $t$ on $\mathcal{X}$. $\mathcal{X}\mathcal{Y}$ is shorthand for $\mathcal{X} \cup \mathcal{Y}$. The empty set is denoted as $\{\}$.
- **Lists.** $\mathbf{X}$, $\mathbf{Y}$ and $\mathbf{Z}$ denote *lists* of attributes. $\mathbf{X}$ may represent the empty list, denoted as $[\,]$. $[\mathsf{A}, \mathsf{B}, \mathsf{C}]$ denotes an explicit list. $[\mathsf{A} \,|\, \mathbf{T}]$ denotes a list with *head* A and *tail* $\mathbf{T}$; i.e., the remaining list when the first element is removed. Let $\mathbf{XY}$ be shorthand for $\mathbf{X} \circ \mathbf{Y}$ ($\mathbf{X}$ concatenate $\mathbf{Y}$). Set $\mathcal{X}$ denotes the *set* of elements in *list* $\mathbf{X}$. Any place a set is expected but a list appears, the list is *cast* to a set; e.g., $t_\mathbf{X}$ denotes $t_\mathcal{X}$. Let $\mathbf{X}'$ denote some other permutation of elements of list $\mathbf{X}$.

Let an *order specification* be a list of attributes defining a lexicographic order; i.e., sorting on the first attribute in the list, breaking ties by the second attribute, breaking further ties by the third attribute, etc. This corresponds to the semantics of the SQL order-by clause.

First, we define the operator '$\preceq_\mathbf{X}$', which defines a *weak total order* over any set of tuples, where $\mathbf{X}$ denotes an order specification. Unless otherwise specified, numbers are ordered numerically, strings are ordered lexicographically and dates are ordered chronologically (all *ascending*).

*Definition 1.* Let $\mathbf{X}$ be a list of attributes. For two tuples $r$ and $s$, $\mathcal{X} \in \mathbf{R}$, $r \preceq_\mathbf{X} s$ if

- $\mathbf{X} = [\,]$; *or*
- $\mathbf{X} = [\mathsf{A} \,|\, \mathbf{T}]$ and $r_\mathsf{A} < s_\mathsf{A}$; or
- $\mathbf{X} = [\mathsf{A} \,|\, \mathbf{T}]$, $r_\mathsf{A} = s_\mathsf{A}$, and $r \preceq_\mathbf{T} s$.

Let $r \prec_\mathbf{X} s$ if $r \preceq_\mathbf{X} s$ but $s \not\preceq_\mathbf{X} r$.

Next, we define order dependencies [13, 22, 25].

*Definition 2.* Let $\mathbf{X}$ and $\mathbf{Y}$ be order specifications, where $\mathcal{X}, \mathcal{Y} \subseteq \mathbf{R}$. $\mathbf{X} \mapsto \mathbf{Y}$ denotes an *order dependency* (OD), read as $\mathbf{X}$ *orders* $\mathbf{Y}$. We write $\mathbf{X} \leftrightarrow \mathbf{Y}$, read as $\mathbf{X}$ and $\mathbf{Y}$ are *order equivalent*, if $\mathbf{X}$ *orders* $\mathbf{Y}$ and $\mathbf{Y}$ *orders* $\mathbf{X}$. Table $\mathbf{r}$ over $\mathbf{R}$ *satisfies* $\mathbf{X} \mapsto \mathbf{Y}$ ($\mathbf{r} \models \mathbf{X} \mapsto \mathbf{Y}$) if, for all $s, t \in \mathbf{r}$, $r \preceq_\mathbf{X} s$ implies $r \preceq_\mathbf{Y} s$. $\mathbf{X} \mapsto \mathbf{Y}$ is said to *hold* for $\mathbf{R}$ ($\mathbf{R} \models \mathbf{X} \mapsto \mathbf{Y}$) if, for each admissible relational instance $\mathbf{r}$ of $\mathbf{R}$, table $\mathbf{r}$ satisfies $\mathbf{X} \mapsto \mathbf{Y}$. $\mathbf{X} \mapsto \mathbf{Y}$ is *trivial* if, for all $\mathbf{r}$, $\mathbf{r} \models \mathbf{X} \mapsto \mathbf{Y}$.

We assume *ascending* (asc) order in the lexicographical ordering, which is the SQL default for any attributes for which directionality is not explicitly indicated. We do not consider *bidirectional* ODs in this paper, which allow a mix of ascending and descending (desc) orders [25]. We also only consider lexicographic order specifications, as per the SQL order-by semantics, and do not consider *pointwise* ODs [7, 8] here. In a pointwise ordering, an instance of a database satisfies a pointwise order dependency $\mathcal{X} \rightsquigarrow \mathcal{Y}$ if, for all tuples $s$ and $t$, for every attribute A in $\mathcal{X}$, $s[\mathsf{A}]$ *op* $t[\mathsf{A}]$ implies that for every attribute B in $\mathcal{Y}$ $s[\mathsf{B}]$ *op* $t[\mathsf{B}]$, where *op* $\epsilon\{<, >, \leq, \geq, =\}$. Working with lexicographical ODs is more useful for query optimization [13, 24, 25] than working with pointwise ODs, because the sequence of the attributes in an order specification as in the order-by statement matters.

Example 1. *Recall Table 1, in which tax is calculated as a percentage of salary, and tax groups and subgroups are based on salary. Tax, percentage and group increase with salary. Furthermore, within the same group, subgroup (*subgroup*) increases with salary. Finally, within the same year, bin increases with salary. Thus, the following ODs hold:* [salary] $\mapsto$ [tax]*;* [salary] $\mapsto$ [percentage]*;* [salary] $\mapsto$ [group, subgroup]*; and* [year, salary] $\mapsto$ [year, bin].

Let Table 1 have a clustered index on year, salary. Then, as discussed earlier, given the OD [year, salary] $\mapsto$ [year, bin], a query with `order by year, bin` could be evaluated using the index on year and salary.

## 2.2 List-based Axiomatization

Figure 1 shows a *sound* and *complete* axiomatization for ODs [22]. The Chain axiom uses the notion of *order compatibility*, denoted as "$\sim$", defined below.

1. *Reflexivity*
$$\mathbf{XY} \mapsto \mathbf{X}$$
2. *Prefix*
$$\frac{\mathbf{X} \mapsto \mathbf{Y}}{\mathbf{ZX} \mapsto \mathbf{ZY}}$$
3. *Transitivity*
$$\frac{\mathbf{X} \mapsto \mathbf{Y} \quad \mathbf{Y} \mapsto \mathbf{Z}}{\mathbf{X} \mapsto \mathbf{Z}}$$
4. *Normalization*
$$\mathbf{WXYXV} \leftrightarrow \mathbf{WXYV}.$$
5. *Suffix*
$$\frac{\mathbf{X} \mapsto \mathbf{Y}}{\mathbf{X} \leftrightarrow \mathbf{YX}}$$
6. *Chain*
$$\frac{\mathbf{X} \sim \mathbf{Y}_1 \quad \forall_{i \in [1,n-1]} \mathbf{Y}_i \sim \mathbf{Y}_{i+1} \quad \mathbf{Y}_n \sim \mathbf{Z} \quad \forall_{i \in [1,n]} \mathbf{Y}_i\mathbf{X} \sim \mathbf{Y}_i\mathbf{Z}}{\mathbf{X} \sim \mathbf{Z}}$$

Figure 1: List-based axioms for ODs.

*Definition 3.* Two order specifications $\mathbf{X}$ and $\mathbf{Y}$ are *order compatible*, denoted as $\mathbf{X} \sim \mathbf{Y}$, *if* $\mathbf{XY} \leftrightarrow \mathbf{YX}$. The empty order specification (i.e., [ ]) is order compatible with *any* order specification.

Example 2. *[d_month] $\sim$ [d_week] is valid: sorting by month and breaking ties by week is equivalent to sorting by week and breaking ties by month. However, the OD [d_month] $\mapsto$ [d_week] does not hold since any given month corresponds to several different weeks (in other words, d_month does not* functionally determine *d_week). Thus, order compatibility is a weaker notion than order dependency.*

## 2.3 Violations

ODs can be violated in two ways. We begin with the following theorem and then explain how the two conditions therein correspond to two possible sources of violations.

Theorem 1. *For every instance* $\mathbf{r}$ *of relation* $\mathbf{R}$*,* $\mathbf{X} \mapsto \mathbf{Y}$ *iff* $\mathbf{X} \mapsto \mathbf{XY}$ *and* $\mathbf{X} \sim \mathbf{Y}$*.*

**Proof**
Suppose $\mathbf{X} \mapsto \mathbf{Y}$. By the *Suffix* rule $\mathbf{X} \leftrightarrow \mathbf{YX}$. Hence, by *Prefix* and *Normalization* $\mathbf{X} \sim \mathbf{Y}$ holds.

The other direction, assume that $\mathbf{X} \mapsto \mathbf{XY}$ and $\mathbf{X} \sim \mathbf{Y}$. By *Transitivity*, $\mathbf{X} \mapsto \mathbf{YX}$. Therefore, by *Reflexivity* and *Transitivity*, $\mathbf{X} \mapsto \mathbf{Y}$. □

There is a strong relationship between ODs and FDs. Any OD implies an FD, modulo lists and sets but not vice versa.

Lemma 1. *For every instance* $\mathbf{r}$ *of relation* $\mathbf{R}$*, if an OD* $\mathbf{X} \mapsto \mathbf{Y}$ *holds, then the FD* $\mathcal{X} \rightarrow \mathcal{Y}$ *is true.*

**Proof**
Let rows $s$ and $t \in \mathbf{r}$. Assume that $s_{\mathcal{X}} = t_{\mathcal{X}}$. Hence, $s \preceq_{\mathbf{X}} t$ and $t \preceq_{\mathbf{X}} s$. By definition of an OD, $s \preceq_{\mathbf{Y}} t$ and $t \preceq_{\mathbf{Y}} s$. Therefore, $s_{\mathcal{Y}} = t_{\mathcal{Y}}$ holds. □

Furthermore, there exists a correspondence between FDs and ODs that is directly related to the first condition in Theorem 1.

Theorem 2. *For relation* $\mathbf{R}$*, for every instance* $\mathbf{r}$*,* $\mathcal{X} \rightarrow \mathcal{Y}$ *iff* $\mathbf{X} \mapsto \mathbf{XY}$*, for any list* $\mathbf{X}$ *over the attributes of* $\mathcal{X}$ *and any list* $\mathbf{Y}$ *over the attributes of* $\mathcal{Y}$*.*

**Proof**
Assume an OD $\mathbf{X} \mapsto \mathbf{XY}$ does not hold. This means, there exist $s$ and $t \in \mathbf{r}$, such that $s \preceq_{\mathbf{X}} t$ but $s \not\preceq_{\mathbf{XY}} t$ by Definition 2. Therefore, $s_{\mathcal{X}} = t_{\mathcal{X}}$ and $s \prec_{\mathbf{Y}} t$. Also $s \prec_{\mathbf{Y}} t$ implies that $s_{\mathcal{Y}} \neq t_{\mathcal{Y}}$. Therefore, $\mathcal{X} \rightarrow \mathcal{Y}$ is not satisfied.

The other direction, by Lemma 1 if $\mathbf{X} \mapsto \mathbf{XY}$, then $\mathcal{X} \rightarrow \mathcal{XY}$. The FD $\mathcal{XY} \rightarrow \mathcal{Y}$ holds by Armstrong's axiom of Reflexivity [2]. Hence, by Armstrong's axiom of Transitivity, $\mathcal{X} \rightarrow \mathcal{Y}$. □

We are now ready to define the two sources of OD violations: *splits* and *swaps.*

*Definition 4.* A *split* with respect to an OD $\mathbf{X} \mapsto \mathbf{XY}$ is a pair of tuples $s$ and $t$ such that $s_{\mathcal{X}} = t_{\mathcal{X}}$ but $s_{\mathcal{Y}} \neq t_{\mathcal{Y}}$. This says that $\mathcal{X}$ does not *functionally determine* $\mathcal{Y}$.

*Definition 5.* A *swap* with respect to $\mathbf{X} \sim \mathbf{Y}$ (i.e., with respect to $\mathbf{XY} \leftrightarrow \mathbf{YX}$) is a pair of tuples $s$ and $t$ such that $s \prec_{\mathbf{X}} t$, but $t \prec_{\mathbf{Y}} s$; thus, the swap falsifies $\mathbf{X} \sim \mathbf{Y}$.

Example 3. *In Table 1, there are three splits with respect to the OD* [position] $\mapsto$ [position, salary] *because* position *does not functionally determine* salary*. The violating tuple pairs are t1 and t4, t2 and t5, and t3 and t6. Also, there is a swap with respect to* [salary] $\sim$ [subgroup]*, e.g., over pair of tuples t1 and t2.*

# 3. SET-BASED CANONICAL FORM

We now present our first set of contributions: a polynomial mapping (from the list-based representation) to a set-based canonical form for ODs, and a *sound* and *complete* axiomatization over this representation. These are the fundamental building blocks of our efficient OD discovery algorithm that will be discussed in Section 4.

## 3.1 Mapping to Canonical Form

Expressing ODs in a natural way relies on lists of attributes, as in the SQL order-by statement. One might well wonder whether lists are inherently necessary. Indeed, we provide a polynomial *mapping* of list-based ODs into *equivalent* set-based canonical ODs. The mapping allows us to develop an efficient OD discovery algorithm that traverses a much smaller set-containment lattice rather than a list-containment lattice used in prior work [13].

Two tuples $s$ and $t$ are *equivalent* with respect to a given set $\mathcal{X}$ if $s_{\mathcal{X}} = t_{\mathcal{X}}$. Any attribute set $\mathcal{X}$ partitions tuples into *equivalence classes.* We denote the *equivalence class* of a tuple $t \in \mathbf{r}$ with respect to a given set $\mathcal{X}$ by $\mathcal{E}(t_{\mathcal{X}})$, i.e., $\mathcal{E}(t_{\mathcal{X}}) = \{s \in \mathbf{r} \mid s_{\mathcal{X}} = t_{\mathcal{X}}\}$. A *partition* of $\mathbf{r}$ over $\mathcal{X}$ is the set of equivalence classes, $\Pi_{\mathcal{X}} = \{\mathcal{E}(t_{\mathcal{X}}) \mid t \in \mathbf{r}\}$. For instance, in Table 1, $\mathcal{E}(t1_{\{\text{year}\}}) = \mathcal{E}(t2_{\{\text{year}\}}) = \mathcal{E}(t3_{\{\text{year}\}}) = \{t1, t2, t3\}$ and $\Pi_{\text{year}} = \{\{t1, t2, t3\}, \{t4, t5, t6\}\}$

We now introduce a *canonical form* for ODs.

*Definition 6.* An attribute A is a *constant* within each equivalence class with respect to $\mathcal{X}$, denoted as $\mathcal{X}$: $[\,] \mapsto$ A, if $\mathbf{X}' \mapsto \mathbf{X}'$A for any permutation $\mathbf{X}'$ of $\mathbf{X}$. Furthermore, two attributes A and B are order-compatible (i.e., no swaps) within each equivalence class with respect to $\mathcal{X}$, denoted as $\mathcal{X}$: A $\sim$ B, if $\mathbf{X}'$A $\sim$ $\mathbf{X}'$B. ODs of the form of $\mathcal{X}$: $[\,] \mapsto$ A and $\mathcal{X}$: A $\sim$ B are called *canonical* ODs, and the set $\mathcal{X}$ is called a *context.*

Example 4. *In Table 1,* bin *is a constant in the context of* position (posit)*, written as* {position}: $[\,] \mapsto$ bin*. This is because* $\mathcal{E}(t1_{\{\text{position}\}}) \models [\,] \mapsto$ bin*,* $\mathcal{E}(t2_{\{\text{position}\}}) \models [\,] \mapsto$ bin *and* $\mathcal{E}(t3_{\{\text{position}\}}) \models [\,] \mapsto$ bin*. Also, there is no swap between* bin *and* salary *in the context of* year*, i.e.,* {year}: bin $\sim$ salary*. This is because* $\mathcal{E}(t1_{\{\text{year}\}}) \models$ bin $\sim$ salary *and* $\mathcal{E}(t4_{\{\text{year}\}}) \models$ bin $\sim$ salary*. However, the canonical ODs* {year}: bin $\sim$ subgroup *and* {position}: $[\,] \mapsto$ salary *do not*

1. Reflexivity
   $\mathcal{X}$: $[\,] \mapsto \mathsf{A}$, $\forall \mathsf{A} \in \mathcal{X}$
2. Identity
   $\mathcal{X}$: $\mathsf{A} \sim \mathsf{A}$
3. Commutativity
   $$\frac{\mathcal{X}\text{: } \mathsf{A} \sim \mathsf{B}}{\mathcal{X}\text{: } \mathsf{B} \sim \mathsf{A}}$$
4. Strengthen
   $$\frac{\mathcal{X}\text{: } [\,] \mapsto \mathsf{A} \quad \mathcal{X}\mathsf{A}\text{: } [\,] \mapsto \mathsf{B}}{\mathcal{X}\text{: } [\,] \mapsto \mathsf{B}}$$
5. Propagate
   $$\frac{\mathcal{X}\text{: } [\,] \mapsto \mathsf{A}}{\mathcal{X}\text{: } \mathsf{A} \sim \mathsf{B}}$$
6. Augmentation-I
   $$\frac{\mathcal{X}\text{: } [\,] \mapsto \mathsf{A}}{\mathcal{Z}\mathcal{X}\text{: } [\,] \mapsto \mathsf{A}}$$
7. Augmentation-II
   $$\frac{\mathcal{X}\text{: } \mathsf{A} \sim \mathsf{B}}{\mathcal{Z}\mathcal{X}\text{: } \mathsf{A} \sim \mathsf{B}}$$
8. Chain
   $$\frac{\mathcal{X}\text{: } \mathsf{A} \sim \mathsf{B}_1 \quad \forall_{i\in[1,n-1]}, \mathcal{X}\text{: } \mathsf{B}_i \sim \mathsf{B}_{i+1} \quad \mathcal{X}\text{: } \mathsf{B}_n \sim \mathsf{C} \quad \forall_{i\in[1,n]}, \mathcal{X}\mathsf{B}_i : \mathsf{A} \sim \mathsf{C}}{\mathcal{X}\text{: } \mathsf{A} \sim \mathsf{C}}$$

Figure 2: Set-based axiomatization for canonical ODs.

*hold as* $\mathcal{E}(t1_{\{\mathsf{year}\}}) \not\models \mathsf{bin} \sim \mathsf{subgroup}$ *and* $\mathcal{E}(t1_{\{\mathsf{position}\}}) \not\models [\,] \mapsto \mathsf{salary}$*, respectively.*

Given a set of attributes $\mathcal{Y}$, for brevity, we state $\forall j$, $\mathsf{Y}_j$ to mean $\forall j \in [1..|\mathcal{Y}|]$, $\mathsf{Y}_j$ in the remaining.

THEOREM 3.
*An OD* $\mathbf{X} \mapsto \mathbf{XY}$ *holds iff* $\forall j$, $\mathcal{X}$: $[\,] \mapsto \mathsf{Y}_j$.

**Proof**
Let $\mathbf{X} \mapsto \mathbf{XY}$ hold. Therefore, by *Reflexivity* and *Transitivity* $\forall j$, $\mathbf{X} \mapsto \mathbf{X}\mathsf{Y}_j$. Hence, by *Prefix* and *Normalization* $\forall j$, $\mathbf{X}' \mapsto \mathbf{X}'\mathsf{Y}_j$. This maps by Definition 6 to $\forall j$, $\mathcal{X}$: $[\,] \mapsto \mathsf{Y}_j$.

To prove the opposite, assume $\forall j$, $\mathcal{X} : [\,] \mapsto \mathsf{Y}_j$. Therefore, $\forall j$, $\mathbf{X}' \mapsto \mathbf{X}'\mathsf{Y}_j$. Hence, it follows from *Union* [22] (if $\mathbf{X} \mapsto \mathbf{Y}$ and $\mathbf{X} \mapsto \mathbf{Z}$, then $\mathbf{X} \mapsto \mathbf{YZ}$) that $\mathbf{X} \mapsto \mathbf{XY}$. □

THEOREM 4.
$\mathbf{X} \sim \mathbf{Y}$ *iff* $\forall i, j$, $\{\mathsf{X}_1, .., \mathsf{X}_{i-1}, \mathsf{Y}_1, .., \mathsf{Y}_{j-1}\}$: $\mathsf{X}_i \sim \mathsf{Y}_j$.

**Proof**
Let $\mathbf{X} \sim \mathbf{Y}$. Therefore, by *Downward Closure* [22] (if $\mathbf{XZ} \sim \mathbf{YV}$, then $\mathbf{XZ} \sim \mathbf{YV}$), $\forall i, j$, $\mathsf{X}_1, .., \mathsf{X}_i \sim \mathsf{Y}_1, .., \mathsf{Y}_j$. Hence, by *Prefix*, *Normalization* and Theorem 1, $\forall i, j$, $\{\mathsf{X}_1, .., \mathsf{X}_{i-1}, \mathsf{Y}_1, .., \mathsf{Y}_{j-1}\}$: $\mathsf{X}_i \sim \mathsf{Y}_j$.

To prove the opposite, given $\forall i, j$, $\{\mathsf{X}_1, .., \mathsf{X}_{i-1}, \mathsf{Y}_1, .., \mathsf{Y}_{j-1}\}$: $\mathsf{X}_i \sim \mathsf{Y}_j$, by *Prefix*, *Normalization*, *Transitivity* and *Reflexivity*, it follows that $\mathbf{X} \sim \mathbf{Y}$. □

Importantly, since $\mathbf{X} \mapsto \mathbf{Y}$ *iff* $\mathbf{X} \mapsto \mathbf{XY}$ and $\mathbf{X} \sim \mathbf{Y}$ (by Theorem 1), by Theorems 3 and 4, a list-based OD can be *mapped* into the following *equivalent* set-based *canonical* ODs. (The size of this mapping is the product $|\mathbf{X}| * |\mathbf{Y}|$.)

THEOREM 5. $\mathbf{X} \mapsto \mathbf{Y}$ *iff* $\forall j$, $\mathcal{X}$: $[\,] \mapsto \mathsf{Y}_j$ *and* $\forall i, j$, $\{\mathsf{X}_1, .., \mathsf{X}_{i-1}, \mathsf{Y}_1, .., \mathsf{Y}_{j-1}\}$: $\mathsf{X}_i \sim \mathsf{Y}_j$. *This provides a polynomial mapping of a list-based OD to equivalent set-based canonical ODs.*

**Proof**
Theorem 1 states that $\mathbf{X} \mapsto \mathbf{Y}$ *iff* $\mathbf{X} \mapsto \mathbf{XY}$ and $\mathbf{X} \sim \mathbf{Y}$. Therefore, by Theorem 3 and 4, $\mathbf{X} \mapsto \mathbf{Y}$ *iff* $\forall j$, $\mathcal{X}$: $[\,] \mapsto \mathsf{Y}_j$ and $\forall i, j$, $\{\mathsf{X}_1, .., \mathsf{X}_{i-1}, \mathsf{Y}_1, .., \mathsf{Y}_{j-1}\}$: $\mathsf{X}_i \sim \mathsf{Y}_j$. Hence, the mapping is polynomial as it is quadratic in the size of an OD $\mathbf{X} \mapsto \mathbf{Y}$; i.e., $|\mathbf{X}| * |\mathbf{Y}|$. □

EXAMPLE 5. *By Theorem 5, an OD* $[\mathsf{AB}] \mapsto [\mathsf{CD}]$ *can be mapped into the following equivalent set of canonical ODs:* $\{\mathsf{A}, \mathsf{B}\}$: $[\,] \mapsto \mathsf{C}$, $\{\mathsf{A}, \mathsf{B}\}$: $[\,] \mapsto \mathsf{D}$, $\{\}$*:* $\mathsf{A} \sim \mathsf{C}$, $\{\mathsf{A}\}$*:* $\mathsf{B} \sim \mathsf{C}$, $\{\mathsf{C}\}$*:* $\mathsf{A} \sim \mathsf{D}$, $\{\mathsf{A}, \mathsf{C}\}$*:* $\mathsf{B} \sim \mathsf{D}$.

## 3.2 Set-based Axiomatization

We present a *sound* and *complete set-based axiomatization* for ODs in Figure 2. For clarity, we denote names of list-based inference rules (Figure 1) with italic font and those of set-based inference rules (Figure 2) with regular font.

THEOREM 6. *The set-based axiomatization for canonical ODs in Figure 2 is sound.*

**Proof**
To prove the soundness, by Theorem 5, it is sufficient to show that all the set-based OD axioms presented in Figure 2 can be inferred from the list-based OD axioms presented in Figure 1.

1. **Reflexivity.**
   It follows from the *Reflexivity* axiom that $\forall \mathsf{A} \in \mathcal{X}$, $\mathbf{X}' \mapsto \mathbf{X}'\mathsf{A}$. Hence, by Theorem 5, $\forall \mathsf{A} \in \mathcal{X}$, $\mathcal{X}$: $[\,] \mapsto \mathsf{A}$.
2. **Identity.**
   It follows from the *Reflexivity* axiom that $\mathbf{X}'\mathsf{A} \mapsto \mathbf{X}'\mathsf{A}$. Therefore, $\mathcal{X}$: $\mathsf{A} \sim \mathsf{A}$ is true by Theorem 5.
3. **Commutativity:** since $\mathcal{X}$: $\mathsf{A} \sim \mathsf{B}$, therefore, $\mathbf{X}'\mathsf{B} \mapsto \mathbf{X}'\mathsf{A}$ (Theorem 5). Hence, $\mathcal{X}$: $\mathsf{B} \sim \mathsf{A}$ is true by Theorem 5.
4. **Strengthen.**
   Given $\mathcal{X}$: $[\,] \mapsto \mathsf{A}$ and $\mathcal{X}\mathsf{A}$: $[\,] \mapsto \mathsf{B}$, $\mathbf{X}' \mapsto \mathbf{X}'\mathsf{A}$ and $\mathbf{X}'\mathsf{A} \mapsto \mathbf{X}'\mathsf{AB}$ hold, respectively (Theorem 5). Therefore, it can be inferred from *Transitivity* that $\mathbf{X}' \mapsto \mathbf{X}'\mathsf{AB}$. Since $\mathbf{X}' \mapsto \mathbf{X}'\mathsf{A}$, it can be deduced from *Suffix*, *Transitivity* and *Normalization* that $\mathbf{X}' \leftrightarrow \mathbf{X}'\mathsf{A}$. Hence, by *Replace* [22] (if $\mathbf{M} \leftrightarrow \mathbf{N}$, then $\mathbf{XMZ} \leftrightarrow \mathbf{XNZ}$) $\mathbf{X}' \mapsto \mathbf{X}'\mathsf{B}$ holds, which can be mapped by Theorem 5 to $\mathcal{X}$: $[\,] \mapsto \mathsf{B}$.
5. **Propagate.**
   Let $\mathcal{X}$: $[\,] \mapsto \mathsf{A}$. Hence, $\mathbf{X}' \mapsto \mathbf{X}'\mathsf{A}$ holds (Theorem 5). It follows by *Suffix*, *Normalization* and *Transitivity* that $\mathbf{X}' \leftrightarrow \mathbf{X}'\mathsf{A}$. By *Reflexivity* $[\mathsf{BA}] \mapsto [\mathsf{B}]$, therefore, by *Prefix* $\mathbf{X}'\mathsf{BA} \mapsto \mathbf{X}'\mathsf{AB}$. Hence, $\mathcal{X}$: $\mathsf{A} \sim \mathsf{B}$ holds (Theorem 5).
6. **Augmentation–I.**
   Let $\mathcal{X}$: $[\,] \mapsto \mathsf{A}$. Therefore, $\mathbf{X}' \mapsto \mathbf{X}'\mathsf{A}$ is true by Theorem 5. By the *Prefix*, *Normalization*, *Reflexivity* and *Transitivity* axioms $[\mathbf{ZX}]' \mapsto [\mathbf{ZX}]'\mathsf{A}$. Hence, $\mathcal{Z}\mathcal{X}$: $[\,] \mapsto \mathsf{A}$ (Theorem 5).
7. **Augmentation–II.**
   Assume $\mathcal{X}$: $[\,] \sim \mathsf{A}$. Therefore, since $\mathbf{X}' \sim \mathbf{X}'\mathsf{A}$ (Theorem 5) by *Normalization* $\mathbf{X}'\mathsf{A} \leftrightarrow \mathbf{X}'\mathsf{A}$. By *Prefix*, *Normalization*, *Reflexivity* and *Transitivity* $\mathbf{ZX}'\mathsf{A} \leftrightarrow \mathbf{ZX}'\mathsf{A}$, which maps by Theorem 5 to $\mathcal{Z}\mathcal{X}$: $[\,] \sim \mathsf{A}$.
8. **Chain.**
   Assume that the ODs $\mathcal{X}$: $\mathsf{A} \sim \mathsf{B}_1$, $\forall_{i\in[1,n-1]}$, $\mathcal{X}$: $\mathsf{B}_i \sim \mathsf{B}_{i+1}$, $\mathcal{X}$: $\mathsf{B}_n \sim \mathsf{C}$ and $\forall_{i\in[1,n]}$, $\mathcal{X}\mathsf{B}_i$: $\mathsf{A} \sim \mathsf{C}$ hold. Therefore, it follows from *Chain* and Definition 6 that $\mathbf{X}'\mathsf{A} \sim \mathbf{X}'\mathsf{C}$. Hence, $\mathcal{X}$: $\mathsf{A} \sim \mathsf{C}$ can be inferred from Theorem 5.

□

Below, we show additional inference rules that can be inferred from the axioms in Figure 2, as they are used in Section 4 and in the proof of completeness (Theorem 7).

LEMMA 2 (TRANSITIVITY).

1. $\forall j,\ \mathcal{X}$: $[\,] \mapsto \mathsf{Y}_j$
2. $\forall k,\ \mathcal{Y}$: $[\,] \mapsto \mathsf{Z}_k$

$\forall k,\ \mathcal{X}$: $[\,] \mapsto \mathsf{Z}_k$

**Proof**

Let $m = |\mathbf{Y}|$

3. $\mathcal{X} \cup \{\mathsf{Y}_1, .., \mathsf{Y}_{m-1}\}$: $[\,] \mapsto \mathsf{Y}_m$ [Aug-I(1)]
4. $\forall k,\ \mathcal{X} \cup \{\mathsf{Y}_1, .., \mathsf{Y}_m\}$: $[\,] \mapsto \mathsf{Z}_k$ [Aug-I(2)]
5. $\forall k,\ \mathcal{X} \cup \{\mathsf{Y}_1, .., \mathsf{Y}_{m-1}\}$: $[\,] \mapsto \mathsf{Z}_k$ [Str(3,4)]
6. $\forall j,\ \mathcal{X} \cup \{\mathsf{Y}_1, .., \mathsf{Y}_{j-1}\}$: $[\,] \mapsto \mathsf{Y}_j$ [Aug-I(1)]
7. $\forall j, k,\ \mathcal{X} \cup \{\mathsf{Y}_1, .., \mathsf{Y}_j\}$: $[\,] \mapsto \mathsf{Z}_k$ [by induction and Strengthen(4-6)]
8. $\forall k,\ \mathcal{X}$ : $[\,] \mapsto \mathsf{Z}_k$ [Strengthen(1,7)]

□

Interestingly (and non-trivially) the set-based Transitivity and Weak Transitivity inference rules correspond to the list-based *Transitivity* axiom.

LEMMA 3 (WEAK TRANS).

1. $\forall i, j,\ \{\mathsf{X}_1, .., \mathsf{X}_{i-1}, \mathsf{Y}_1, .., \mathsf{Y}_{j-1}\}$: $\mathsf{X}_i \sim \mathsf{Y}_j$
2. $\forall j, k,\ \{\mathsf{Y}_1, .., \mathsf{Y}_{j-1}, \mathsf{Z}_1, .., \mathsf{Z}_{k-1}\}$: $\mathsf{Y}_j \sim \mathsf{Z}_k$
3. $\forall k,\ \mathcal{Y}$: $[\,] \mapsto \mathsf{Z}_k$

$\forall i, k,\ \{\mathsf{X}_1, .., \mathsf{X}_{i-1}, \mathsf{Z}_1, .., \mathsf{Z}_{k-1}\}$: $\mathsf{X}_i \sim \mathsf{Z}_k$

**Proof**

Let $m = |\mathbf{Y}|$

4. $\forall i,\ \{\mathsf{X}_1, .., \mathsf{X}_{i-1}, \mathsf{Y}_1, .., \mathsf{Y}_m\}$: $\mathsf{X}_i \sim \mathsf{Y}_m$ [Given(1)]
5. $\forall k,\ \{\mathsf{Y}_1, .., \mathsf{Y}_m, \mathsf{Z}_1, .., \mathsf{Z}_{k-1}\}$: $\mathsf{Y}_m \sim \mathsf{Z}_k$ [Given(2)]
6. $\forall i, k,\ \{\mathsf{Y}_1, .., \mathsf{Y}_m\}$: $\mathsf{X}_i \sim \mathsf{Z}_k$ [Propagation(3)]
7. $\forall i, k,\ \{\mathsf{X}_1, .., \mathsf{X}_{i-1}, \mathsf{Z}_1, .., \mathsf{Z}_{k-1}, \mathsf{Y}_1, .., \mathsf{Y}_m\}$: $\mathsf{X}_i \sim \mathsf{Z}_k$ [Chain, Aug–II(4-6)]
8. $\forall i, k, j,\ \{\mathsf{X}_1, .., \mathsf{X}_{i-1}, \mathsf{Z}_1, .., \mathsf{Z}_{k-1}, \mathsf{Y}_1, .., \mathsf{Y}_j\}$: $\mathsf{X}_i \sim \mathsf{Z}_k$ [by induction; Chain, Aug–II(1,2,7)]
9. $\forall i, k,\ \{\mathsf{X}_1, .., \mathsf{X}_{i-1}, \mathsf{Z}_1, .., \mathsf{Z}_{k-1}$: $\mathsf{X}_i \sim \mathsf{Z}_k$ [Chain, Aug–II(1,2,8)]

□

LEMMA 4 (NORMALIZATION).

1. $\mathcal{X}$: $\mathsf{A} \sim \mathsf{B},\ \forall \mathsf{A} \in \mathcal{X}$

**Proof**

2. $\forall i,\ \mathcal{X}$ : $[\,] \mapsto \mathsf{X}_i$ [Reflexivity]
3. $\forall i,\ \mathcal{X}$ : $[\,] \mapsto \mathsf{X}_i$ [Prop(2)]

□

The Normalization inference rule allows us to eliminate *trivial* dependencies over swaps.

THEOREM 7. *The proposed set-based axiomatization for canonical ODs in Figure 2 is sound and complete.*

**Proof**

Since the set-based OD axioms are sound (Theorem 6), the remaining step is to show the completeness, i.e, that all the list-based OD axioms follow from the set-based OD axioms.

1. ***Reflexivity***.
$\mathbf{XY} \mapsto \mathbf{X}$ *iff* $\forall i,\ \mathcal{XY}$: $[\,] \mapsto \mathsf{X}_i$ and $\forall i,\ \{\mathsf{X}_1, .., \mathsf{X}_{i-1}\}$: $\mathsf{X}_i \sim \mathsf{X}_i$ by Theorem 5. However, the OD $\mathcal{XY}$: $[\,] \mapsto \mathsf{X}_i$ follows from Reflexivity and $\forall i,\ \{\mathsf{X}_1, .., \mathsf{X}_{i-1}\}$: $\mathsf{X}_i \sim \mathsf{X}_i$ can be inferred from the Identity axiom.
2. ***Prefix***.
Assume that $\mathbf{X} \mapsto \mathbf{Y}$ holds. $\mathbf{X} \mapsto \mathbf{Y}$ is equivalent by Theorem 5 to $\forall j,\ \mathcal{X}$: $[\,] \mapsto \mathsf{Y}_j$ and $\forall i, j,\ \{\mathsf{X}_1, .., \mathsf{X}_{i-1}, \mathsf{Y}_1, .., \mathsf{Y}_{j-1}\}$: $\mathsf{X}_i \sim \mathsf{Y}_j$. Therefore, it follows by Augmentation–I that $\forall j,\ \mathcal{ZX}$: $[\,] \mapsto \mathsf{Y}_j$. By Reflexivity, it can be inferred that $\forall k,\ \mathcal{ZX}$: $[\,] \mapsto \mathsf{Z}_k$. Since, $\forall i, j,\ \{\mathsf{X}_1, .., \mathsf{X}_{i-1}, \mathsf{Y}_1, .., \mathsf{Y}_{j-1}\}$: $\mathsf{X}_i \sim \mathsf{Y}_j$, by Augmentation–II, it can be inferred that $\forall i, j,\ \{\mathcal{Z} \cup \mathsf{X}_1, .., \mathsf{X}_{i-1}, \mathsf{Y}_1, .., \mathsf{Y}_{j-1}\}$: $\mathsf{X}_i \sim \mathsf{Y}_j$. Additionally it follows from Identity that $\forall k,\ \{\mathsf{Z}_1, .., \mathsf{Z}_{k-1}\}$: $\mathsf{Z}_k \sim \mathsf{Z}_k$. Therefore, by Theorem 5, it follows that the OD $\mathbf{ZX} \mapsto \mathbf{ZY}$ is true.
3. ***Transitivity***.
Let $\mathbf{X} \mapsto \mathbf{Y}$ and $\mathbf{Y} \mapsto \mathbf{Z}$. $\mathbf{X} \mapsto \mathbf{Y}$ is equivalent to $\forall j,\ \mathcal{X}$: $[\,] \mapsto \mathsf{Y}_j$ and $\forall i, j,\ \{\mathsf{X}_1, .., \mathsf{X}_{i-1}, \mathsf{Y}_1, .., \mathsf{Y}_{j-1}\}$: $\mathsf{X}_i \sim \mathsf{Y}_j$(Theorem 5). $\mathbf{Y} \mapsto \mathbf{Z}$ is equivalent to $\forall k,\ \mathcal{Y}$: $[\,] \mapsto \mathsf{Z}_k$ and $\forall j, k,\ \{\mathsf{Y}_1, .., \mathsf{Y}_{j-1}, \mathsf{Z}_1, .., \mathsf{Z}_{k-1}\}$: $\mathsf{Y}_j \sim \mathsf{Z}_k$. Hence, by Transitivity (Theorem 2), $\forall k,\ \mathcal{X}$: $[\,] \mapsto \mathsf{Z}_k$ and by Weak Transitivity (Theorem 2) $\forall i, k,\ \{\mathsf{X}_1, .., \mathsf{X}_{i-1}, \mathsf{Z}_1, .., \mathsf{Z}_{k-1}\}$: $\mathsf{X}_i \sim \mathsf{Y}_k$. This is equivalent by Theorem 5 to $\mathbf{X} \mapsto \mathbf{Y}$.
4. ***Suffix***.
Assume $\mathbf{X} \mapsto \mathbf{Y}$ is true. Therefore, the following canonical rule $\forall j,\ \mathcal{X}$: $[\,] \mapsto \mathsf{Y}_j$ is given by Theorem 5. It follows from Reflexivity that $\forall i,\ \mathcal{X}$: $[\,] \mapsto \mathsf{X}_i$ and $\forall i,\ \mathcal{YX}$: $[\,] \mapsto \mathsf{X}_i$. Furthermore, $\forall i, j,\ \{\mathsf{X}_1, .., \mathsf{X}_{i-1}, \mathsf{Y}_1, .., \mathsf{Y}_{j-1}\}$: $\mathsf{X}_i \sim \mathsf{Y}_j$ is given by Theorem 5 as $\mathbf{X} \mapsto \mathbf{Y}$. Hence, $\mathbf{X} \leftrightarrow \mathbf{YX}$ by Theorem 5.
5. ***Normalization***. Let $\mathbf{Z} = \mathbf{XYM}$. It follows from Reflexivity that $\forall k,\ \mathcal{ZW}$: $[\,] \mapsto \mathsf{Z}_k$ and $\forall l,\ \mathcal{ZW}$: $[\,] \mapsto \mathsf{W}_l$. Furthermore, it can be inferred from Identity that $\forall k,\ \{\mathsf{Z}_1, .., \mathsf{Z}_{k-1}\}$: $\mathsf{Z}_k \sim \mathsf{Z}_k$ and $\forall p,\ \mathcal{Z} \cup \{\mathsf{W}_1, .., \mathsf{W}_{p-1}\}$: $\mathsf{W}_p \sim \mathsf{W}_p$. It follows from Reflexivity, Normalization and Augmentation that $\forall i, p,\ \mathcal{Z} \cup \{\mathsf{W}_1, .., \mathsf{W}_{p-1}\}$: $\mathsf{X}_i \sim \mathsf{W}_p$. Therefore, $\mathbf{XYMW} \leftrightarrow \mathbf{XYMYW}$ holds by Theorem 5.
6. ***Chain***.
It follows from the Chain axiom and by mapping $\mathbf{X} \sim \mathbf{Y}_1$, $\forall_{i \in [1,n-1]}, \mathbf{Y}_i \sim \mathbf{Y}_{i+1}$, $\mathbf{Y}_n \sim \mathbf{Z}$ and $\forall_{i \in [1,n]}, \mathbf{Y}_i\mathbf{X} \sim \mathbf{Y}_i\mathbf{Z}$ into the canonical form rules (Theorem 5) that $\forall i, k,\ \{\mathsf{X}_1, .., \mathsf{X}_i, \mathsf{Z}_1, .., \mathsf{Z}_k\}$: $\mathsf{X}_i \sim \mathsf{Z}_k$. This is equivalent by Theorem 5 to $\mathbf{X} \sim \mathbf{Z}$.

This ends the proof of soundness and completeness of set-based OD axiomatization. □

EXAMPLE 6. *Assume* $\{\mathsf{salary}\}$: $[\,] \mapsto \mathsf{tax}$ *as in Table 1. Hence, it can be inferred from the Propagate axiom that* $\{\mathsf{salary}\}$: $\mathsf{tax} \sim \mathsf{year}$.

By Theorem 7, three set-based axioms, namely Reflexivity, Strengthen and Augmentation–I, are sound and complete for the FD fragment of the OD class, as these are the only canonical OD axioms that use the constant operator. In contrast, the sound and complete list-based axiomatization for ODs [22] is concealed within the first five axioms in Figure 1. The versatility and separability of the set-based axioms–between the FD fragment and the order-compatible fragment–allows us to design effective pruning rules for our OD discovery algorithm (Section 4).

# 4. OD DISCOVERY ALGORITHM

## 4.1 FASTOD Main Algorithm

Given the mapping of a list-based OD into the equivalent set-based ODs (Section 3.1), we present an algorithm, named *FASTOD*, which *efficiently* discovers a complete, minimal set of set-based ODs over a given relation instance.

A canonical OD $\mathcal{X}$: $[\,] \mapsto \mathsf{A}$ is *trivial* if $\mathsf{A} \in \mathcal{X}$ (Reflexivity). A canonical OD $\mathcal{X}$: $\mathsf{A} \sim \mathsf{B}$ is *trivial* if $\mathsf{A} \in \mathcal{X}$ or $\mathsf{B} \in \mathcal{X}$ (Normalization, Lemma 4) or $\mathsf{A} = \mathsf{B}$ (Identity). A canonical OD $\mathcal{X}$: $[\,] \mapsto \mathsf{A}$ is *minimal* if it is non-trivial and there is no context $\mathcal{Y} \subset \mathcal{X}$ such that $\mathcal{Y}$: $[\,] \mapsto \mathsf{A}$ holds in **r** (Augmentation–I). A canonical OD $\mathcal{X}$: $\mathsf{A} \sim \mathsf{B}$ is *minimal* if it is non-trivial and there is no context $\mathcal{Y}$, where $\mathcal{Y} \subset \mathcal{X}$, such that $\mathcal{Y}$: $\mathsf{A} \sim \mathsf{B}$ holds in $\mathbf{r}$ (Augmentation–II), or $\mathcal{X}$: $[\,] \mapsto \mathsf{A}$ or $\mathcal{X}$: $[\,] \mapsto \mathsf{B}$ (Propagate) holds in $\mathbf{r}$. Our goal is to compute a complete, minimal set of ODs that hold in $\mathbf{r}$.

*FASTOD* traverses a *lattice* of all possible *sets* of attributes in a level-wise manner (Figure 3) since list-based ODs can be mapped into the equivalent set-based ODs (Theorem 5). In contrast, the OD discovery algorithm from [13] traverses a lattice of all possible *lists* of attributes, which leads to factorial time complexity. In level $\mathcal{L}_l$, our algorithm generates candidate ODs with $l$ attributes using *computeODs*($\mathcal{L}_l$). *FASTOD* starts the search from singleton sets of attributes and works its way to larger attribute sets through the set-containment lattice, level by level. When the algorithm is processing an attribute set $\mathcal{X}$, it verifies ODs of the form $\mathcal{X} \setminus \mathsf{A}$: $[\,] \mapsto \mathsf{A}$ (let $\mathcal{X} \setminus \mathsf{A}$ be shorthand for $\mathcal{X} \setminus \{\mathsf{A}\}$), where $\mathsf{A} \in \mathcal{X}$ and $\mathcal{X} \setminus \{\mathsf{A}, \mathsf{B}\}$: $\mathsf{A} \sim \mathsf{B}$, where $\mathsf{A}$, $\mathsf{B} \in \mathcal{X}$ and $\mathsf{A} \neq \mathsf{B}$. This guarantees that only non-trivial ODs are considered.

**Algorithm 1** FASTOD

**Input:** Relation $\mathbf{r}$ over schema $\mathbf{R}$
**Output:** Minimal set of ODs $\mathcal{M}$, such that $\mathbf{r} \models \mathcal{M}$

1: $\mathcal{L}_0 = \{\}$
2: $\mathcal{C}_c^+(\{\}) = \mathbf{R}$
3: $\mathcal{C}_s^+(\{\}) = \{\}$
4: $l = 1$
5: $\mathcal{L}_1 = \{\mathsf{A} \mid \mathsf{A} \in \mathbf{R}\}$
6: $\forall_{\mathsf{A} \in R} \mathcal{C}_s^+(\mathsf{A}) = \{\}$
7: **while** $\mathcal{L}_l \neq \{\}$ **do**
8:     *computeODs*($\mathcal{L}_l$)
9:     *pruneLevels*($\mathcal{L}_l$)
10:     $\mathcal{L}_{l+1}$ = *calculateNextLevel*($\mathcal{L}_l$)
11:     $l = l + 1$
12: **end while**
13: **return** $\mathcal{M}$

The small-to-large search strategy of the discovery algorithm guarantees that only ODs that are minimal with respect to the context are added to the output set of ODs $\mathcal{M}$, and is used to prune the search space effectively. The OD candidates generated in a given level are checked for *minimality* based on the previous levels and are added to a *valid* set of ODs $\mathcal{M}$ if applicable. The algorithm *pruneLevels*($\mathcal{L}_l$) prunes the search space by deleting sets from $\mathcal{L}_l$ with the knowledge gained during the checks in *computeODs*($\mathcal{L}_l$). The algorithm *calculateNextLevel*($\mathcal{L}_l$) forms the next level from the current level.

Our algorithm, for example, can detect the following ODs in the TPC-DS benchmark: {d_date_sk}: $[\,] \mapsto$ [d_date]; {}: [d_date_sk] $\sim$ [d_date]; {d_date_sk}: $[\,] \mapsto$ [d_year]; {}: [d_date_sk] $\sim$ [d_year]; {d_month}: $[\,] \mapsto$ [d_quarter] and; {}: [d_month] $\sim$ [d_quarter]. These types of canonical ODs have been proven to be useful to eliminate joins and simplify group-by and order-by statements [23, 25].

Next, we explain, in turn, each of the algorithms that are called in the main loop of *FASTOD* (Lines 7–12).

## 4.2 Finding Minimal ODs

*FASTOD* traverses the lattice until complete, minimal ODs that are valid are found. First, we deal with ODs of the form $\mathcal{X} \setminus \mathsf{A}$: $[\,] \mapsto \mathsf{A}$, where $\mathsf{A} \in \mathcal{X}$. To check if such a OD is minimal, we need to know if $\mathcal{Y} \setminus \mathsf{A}$: $[\,] \mapsto \mathsf{A}$ is valid for $\mathcal{Y} \subset \mathcal{X}$. If $\mathcal{Y} \setminus \mathsf{A}$: $[\,] \mapsto \mathsf{A}$, then by Augmentation–I $\mathcal{X} \setminus \mathsf{A}$: $[\,] \mapsto \mathsf{A}$ holds. An OD $\mathcal{X}$: $[\,] \mapsto \mathsf{A}$ holds for any relational instance by Reflexivity, therefore, considering only $\mathcal{X} \setminus \mathsf{A}$: $[\,] \mapsto \mathsf{A}$ guarantees that only non-trivial ODs are taken into account.

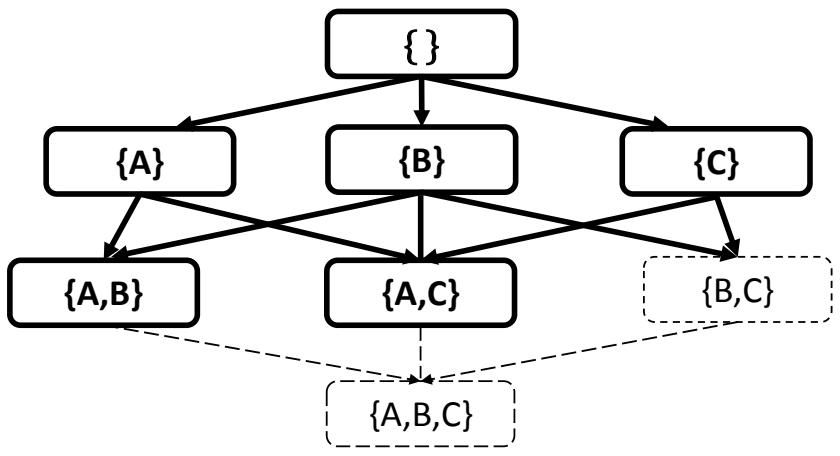

Figure 3: A pruned set lattice for attributes A, B, C. As {B, C} is discarded, only the non-dashed bold parts are accessed.

We maintain information about minimal ODs, in the form of $\mathcal{X} \setminus \mathsf{A}$: $[\,] \mapsto \mathsf{A}$, in the *candidate* set $\mathcal{C}_c^+(\mathcal{X})$ [11].[2] If $\mathsf{A} \in \mathcal{C}_c^+(\mathcal{X})$ for a given set $\mathcal{X}$, then $\mathsf{A}$ has not been found to depend on any proper subset of $\mathcal{X}$. Therefore, to find minimal ODs, it suffices to verify ODs $\mathcal{X} \setminus \mathsf{A}$: $[\,] \mapsto \mathsf{A}$, where $\mathsf{A} \in \mathcal{X}$ and $\mathsf{A} \in \mathcal{C}_c^+(\mathcal{X} \setminus \mathsf{B})$ for all $\mathsf{B} \in \mathcal{X}$.

EXAMPLE 7. *Assume that* $\{\mathsf{B}\}$: $[\,] \mapsto \mathsf{A}$ *and that we consider the set* $\mathcal{X} = \{\mathsf{A}, \mathsf{B}, \mathsf{C}\}$. *As* $\{\mathsf{B}\}$: $[\,] \mapsto \mathsf{A}$ *holds,* $\mathsf{A} \notin \mathcal{C}_c^+(\mathcal{X} \setminus \mathsf{C})$. *Hence, the OD* $\{\mathsf{B}, \mathsf{C}\}$: $[\,] \mapsto \mathsf{A}$ *is not minimal.*

We now show how to prune the search space more effectively. Assume that $\mathsf{B} \in \mathcal{X}$ and an OD $\mathcal{X} \setminus \mathsf{B}$: $[\,] \mapsto \mathsf{B}$ holds. Then, by Lemma 5 below, inferring via the Strengthen axiom, the OD $\mathcal{X}$: $[\,] \mapsto \mathsf{A}$ cannot be minimal because $\mathsf{B}$ can be removed from the context $\mathcal{X}$. Observe that *FASTOD* does not have to know whether $\mathcal{X}$: $[\,] \mapsto \mathsf{A}$ holds.

LEMMA 5. *Let* $\mathsf{B} \in \mathcal{X}$ *and* $\mathcal{X} \setminus \mathsf{B}$*:* $[\,] \mapsto \mathsf{B}$. *If* $\mathcal{X}$: $[\,] \mapsto \mathsf{A}$, *then* $\mathcal{X} \setminus \mathsf{B}$: $[\,] \mapsto \mathsf{A}$.

Hence, we define the candidate set $\mathcal{C}_c^+(\mathcal{X})$, formally as follows. (Note that $\mathsf{A}$ may equal $\mathsf{B}$ in Definition 7).

*Definition 7.* $\mathcal{C}_c^+(\mathcal{X}) = \{\mathsf{A} \in \mathbf{R} \mid$
$\forall_{\mathsf{B} \in \mathcal{X}}\ \mathcal{X} \setminus \{\mathsf{A}, \mathsf{B}\}$: $[\,] \mapsto \mathsf{B}$ does not hold$\}$

EXAMPLE 8. *Assume that* $\{\mathsf{B}\}$*:* $[\,] \mapsto \mathsf{C}$ *and that* FASTOD *considers* $\{\mathsf{B}, \mathsf{C}\}$*:* $[\,] \mapsto \mathsf{A}$. *Since* $\{\mathsf{B}\}$*:* $[\,] \mapsto \mathsf{C}$ *holds, the OD* $\{\mathsf{B}, \mathsf{C}\}$*:* $[\,] \mapsto \mathsf{A}$ *is not minimal and* $\mathsf{A} \notin \mathcal{C}_c^+(\mathcal{X})$, *where* $\mathcal{X} = \{\mathsf{A}, \mathsf{B}, \mathsf{C}\}$.

We now deal with ODs involving '$\sim$'. To verify if a potential OD of the form $\mathcal{X} \setminus \{\mathsf{A}, \mathsf{B}\}$: $\mathsf{A} \sim \mathsf{B}$, where $\mathsf{A}, \mathsf{B} \in \mathcal{X}$ and $\mathsf{A} \neq \mathsf{B}$, is *minimal*, we need to know if $\mathcal{Y} \setminus \{\mathsf{A}, \mathsf{B}\}$: $\mathsf{A} \sim \mathsf{B}$ holds for some proper subset $\mathcal{Y}$ of $\mathcal{X} \setminus \{\mathsf{A}, \mathsf{B}\}$, and that $\mathcal{X} \setminus \{\mathsf{A}, \mathsf{B}\}$: $[\,] \mapsto \mathsf{A}$ and $\mathcal{X} \setminus \{\mathsf{A}, \mathsf{B}\}$: $[\,] \mapsto \mathsf{B}$ do not hold. If $\mathcal{Y} \setminus \{\mathsf{A}, \mathsf{B}\}$: $\mathsf{A} \sim \mathsf{B}$, then by Augmentation–II, $\mathcal{X} \setminus \{\mathsf{A}, \mathsf{B}\}$: $\mathsf{A} \sim \mathsf{B}$ holds. Also, if $\mathcal{X} \setminus \{\mathsf{A}, \mathsf{B}\}$: $[\,] \mapsto \mathsf{A}$ or $\mathcal{X} \setminus \{\mathsf{A}, \mathsf{B}\}$: $[\,] \mapsto \mathsf{B}$, then by Propagate, $\mathcal{X} \setminus \{\mathsf{A}, \mathsf{B}\}$: $\mathsf{A} \sim \mathsf{B}$ holds. ODs $\mathcal{X} \setminus \mathsf{B}$: $\mathsf{A} \sim \mathsf{B}$, $\mathcal{X} \setminus \mathsf{A}$: $\mathsf{A} \sim \mathsf{B}$, and OD $\mathcal{X}$: $\mathsf{A} \sim \mathsf{A}$ are always true by Normalization and Identity, respectively. Hence, considering ODs in the form of $\mathcal{X} \setminus \{A, B\}$: $A \sim B$ guarantees that non-trivial ODs are taken into account.

[2]Since FDs are subsumed by ODs, FD discovery is part of *FASTOD*, and some of our techniques are similar to those from *TANE* [11]. However, *FASTOD* and *TANE* differ in many details even for FD discovery; e.g., a pruning rule for removing nodes from the lattice (Section 4.5) and a key pruning rule (Section 4.6). Additionally, *FASTOD* includes new OD-specific pruning rules.

We store information about minimal ODs in the form of $\mathcal{X} \setminus \{A, B\}$: $A \sim B$, in the *candidate* set $\mathcal{C}_s^+(\mathcal{X})$. If $\{A, B\} \in \mathcal{C}_s^+(\mathcal{X})$ for a given set $\mathcal{X}$, then $A \sim B$ has not been found to hold within the context of any subset of $\mathcal{X} \setminus \{A, B\}$; also, $\mathcal{X} \setminus \{A, B\}$: $[] \mapsto A$ and $\mathcal{X} \setminus \{A, B\}$: $[] \mapsto B$ do not hold. By Commutativity, if $\mathcal{X} \setminus \{A, B\}$: $A \sim B$, then $\mathcal{X} \setminus \{A, B\}$: $B \sim A$. Hence, only $\{A, B\}$ is stored in $\mathcal{C}_s^+(\mathcal{X})$ instead of both $[A, B]$ and $[B, A]$, since order compatibility is symmetric.

EXAMPLE 9. *Let* $\{\}$: $A \sim B$, $\{A\}$: $[] \mapsto C$ *and* $\mathcal{X} = \{A, B, C\}$. *Consider that* $\{C\}$: $A \sim B$ *and* $\{A\}$: $B \sim C$. *Hence,* $\{C\}$: $A \sim B$ *is not minimal as* $\{\}$: $A \sim B$*; therefore,* $\{A, B\} \notin \mathcal{C}_s^+(\mathcal{X})$. *Also,* $\{A\}$: $B \sim C$ *is not minimal since* $\{A\}$: $[] \mapsto C$*; therefore,* $\{B, C\} \notin \mathcal{C}_s^+(\mathcal{X})$.

Again, it is possible to prune the search space more effectively. Assume that for some $C \in \mathcal{X}$, an OD $\mathcal{X} \setminus C$: $[] \mapsto C$ holds. Therefore, by Lemma 6, inferred from Propagate and Chain, the OD $\mathcal{X}$: $A \sim B$ could not be minimal as C can be removed from the context $\mathcal{X}$. Note that the *FASTOD* algorithm does not have to know whether $\mathcal{X}$: $A \sim B$ holds.

LEMMA 6. *Let* $C \in \mathcal{X}$ *and* $\mathcal{X} \setminus C$: $[] \mapsto C$ *hold. If* $\mathcal{X}$: $A \sim B$, *then* $\mathcal{X} \setminus C$: $A \sim B$.

**Proof**

Since $\mathcal{X} \setminus C$: $[] \mapsto C$, therefore, by Propagate $\mathcal{X} \setminus C$: $A \sim C$ and $\mathcal{X} \setminus C$: $B \sim C$. Hence, as $\mathcal{X}$: $A \sim B$ is given, $\mathcal{X} \setminus C$: $A \sim B$ follows from Chain. □

As a result, we define the candidate set $\mathcal{C}_s^+(\mathcal{X})$ as follows. Note that C can be equal to A or B.

*Definition 8.* $\mathcal{C}_s^+(\mathcal{X}) = \{\{A, B\} \in \mathcal{X}^2 \mid A \neq B$ and $\forall_{C \in \mathcal{X}}$ $\mathcal{X} \setminus \{A, B, C\}$: $A \sim B$ does not hold, and $\forall_{C \in \mathcal{X}}$ $\mathcal{X} \setminus \{A, B, C\}$: $[] \mapsto C$ does not hold$\}$

EXAMPLE 10. *Assume that* $\{C\}$: $[] \mapsto D$ *and that* FASTOD *considers* $\{C, D\}$: $A \sim B$. *Since* $\{C\}$: $[] \mapsto D$, *the OD* $\{C, D\}$: $A \sim B$ *is not minimal, and therefore,* $\{A, B\} \notin \mathcal{C}_s^+(\mathcal{X})$, *where* $\mathcal{X} = \{A, B, C, D\}$.

*FASTOD*'s candidate sets do not increase in size during the execution of the algorithm (unlike *ORDER* [13]) because of the concise candidate representation. Also, we show in Section 5.3 than many "minimal" ODs in [13] are considered non-minimal (redundant) by our algorithm. Additional optimizations for our OD discovery algorithm are described in Sections 4.5 and 4.6.

## 4.3 Computing Levels

Algorithm 2 explains *calculateNextLevel*, which computes $\mathcal{L}_{l+1}$ from $\mathcal{L}_l$. It uses the subroutine *singleAttrDifferBlocks*($\mathcal{L}_l$) that partitions $\mathcal{L}_l$ into blocks (Line 2). Two sets belong to the same block if they have a common subset $\mathcal{Y}$ of length $l-1$ and differ in only one attribute, A and B, respectively. Therefore, the blocks are not difficult to calculate as sets $\mathcal{Y}A$ and $\mathcal{Y}B$ can be preserved as sorted sets of attributes. Unlike our *FASTOD* algorithm and other typical uses of Apriori [1], the OD discovery algorithm from [13] generates *all permutations* of attributes of size $l+1$ with the same *prefix* blocks of size $l-1$, which leads to its *factorial* worst-case time complexity.

The level $\mathcal{L}_{l+1}$ contains only those sets of attributes of size $l+1$ which have all their subsets of size $l$ in $\mathcal{L}_l$ (Line 4).

**Algorithm 2** calculateNextLevel($\mathcal{L}_l$)

1: $\mathcal{L}_l = []$
2: **for all** $\{\mathcal{Y}B, \mathcal{Y}C\} \in$ *singleAttrDiffBlocks*($\mathcal{L}_l$) **do**
3: $\quad \mathcal{X} = \mathcal{Y} \bigcup \{B, C\}$
4: $\quad$ **if** $\forall_{A \in \mathcal{X}}\ \mathcal{X} \setminus A \in \mathcal{L}_l$ **then**
5: $\quad\quad$ Add $\mathcal{X}$ to $\mathcal{L}_{l+1}$
6: $\quad$ **end if**
7: **end for**
8: **return** $\mathcal{L}_{l+1}$

## 4.4 Computing Dependencies

Algorithm 3 shows *computeODs*($\mathcal{L}_l$), which adds minimal ODs from level $\mathcal{L}_l$ to $\mathcal{M}$, in the form of $\mathcal{X} \setminus A$: $[] \mapsto A$ and $\mathcal{X} \setminus \{A, B\}$: $A \sim B$, where $A, B \in \mathcal{X}$ and $A \neq B$. The following lemma shows that we can use the candidate set $\mathcal{C}_c^+(\mathcal{X})$ to test whether $\mathcal{X} \setminus A$: $[] \mapsto A$ is minimal [11, 10].

LEMMA 7. *An OD* $\mathcal{X} \setminus A$: $[] \mapsto A$, *where* $A \in \mathcal{X}$, *is minimal iff* $\forall_{B \in \mathcal{X}} A \in \mathcal{C}_c^+(\mathcal{X} \setminus B)$.

**Algorithm 3** computeODs($\mathcal{L}_l$)

**for all** $\mathcal{X} \in \mathcal{L}_l$ **do**
$\quad \mathcal{C}_c^+(\mathcal{X}) = \bigcap_{A \in \mathcal{X}} \mathcal{C}_c^+(\mathcal{X} \setminus A)$
$\quad$ **if** $l = 2$ **then**
$\quad\quad \forall_{A,B \in \mathbf{R}^2, A \neq B} \mathcal{C}_s^+(\{A, B\}) = \{A, B\}$
$\quad$ **else if** $l > 2$ **then**
$\quad\quad \mathcal{C}_s^+(\mathcal{X}) = \{\{A, B\} \in \bigcup_{C \in \mathcal{X}} \mathcal{C}_s^+(\mathcal{X} \setminus C) \mid$
$\quad\quad \forall_{D \in \mathcal{X} \setminus \{A,B\}} \{A, B\} \in \mathcal{C}_s^+(\mathcal{X} \setminus D)\}$
$\quad$ **end if**
**end for**
**for all** $\mathcal{X} \in \mathcal{L}_l$ **do**
$\quad$ **for all** $A \in \mathcal{X} \cap \mathcal{C}_c^+(\mathcal{X})$ **do**
$\quad\quad$ **if** $\mathcal{X} \setminus A$: $[] \mapsto A$ **then**
$\quad\quad\quad$ Add $\mathcal{X} \setminus A$: $[] \mapsto A$ to $\mathcal{M}$
$\quad\quad\quad$ Remove A from $\mathcal{C}_c^+(\mathcal{X})$
$\quad\quad\quad$ Remove all $B \in \mathbf{R} \setminus \mathcal{X}$ from $\mathcal{C}_c^+(\mathcal{X})$
$\quad\quad$ **end if**
$\quad$ **end for**
$\quad$ **for all** $\{A, B\} \in \mathcal{C}_s^+(\mathcal{X})$ **do**
$\quad\quad$ **if** $A \notin \mathcal{C}_c^+(\mathcal{X} \setminus B)$ or $B \notin \mathcal{C}_c^+(\mathcal{X} \setminus A)$ **then**
$\quad\quad\quad$ Remove $\{A, B\}$ from $\mathcal{C}_s^+(\mathcal{X})$
$\quad\quad$ **else if** $\mathcal{X} \setminus \{A, B\}$: $A \sim B$ **then**
$\quad\quad\quad$ Add $\mathcal{X} \setminus \{A, B\}$: $A \sim B$ to $\mathcal{M}$
$\quad\quad\quad$ Remove $\{A, B\}$ from $\mathcal{C}_s^+(\mathcal{X})$
$\quad\quad$ **end if**
$\quad$ **end for**
**end for**

Similarly, we prove in Lemma 8 that we can use the candidate set $\mathcal{C}_s^+(\mathcal{X})$ to verify if $\mathcal{X} \setminus \{A, B\}$: $A \sim B$ is minimal.

LEMMA 8. *The OD* $\mathcal{X} \setminus \{A, B\}$: $A \sim B$, *where* A, $B \in \mathcal{X}$ *and* $A \neq B$, *is minimal iff* $\forall_{C \in \mathcal{X} \setminus \{A,B\}}$ $\{A, B\} \in \mathcal{C}_s^+(\mathcal{X} \setminus C)$, *and* $A \in \mathcal{C}_c^+(\mathcal{X} \setminus B)$ *and* $B \in \mathcal{C}_c^+(\mathcal{X} \setminus A)$.

**Proof**

Assume first that $\mathcal{X} \setminus \{A, B\}$: $A \sim B$ is not minimal. Therefore, there exists $C \in \mathcal{X} \setminus \{A, B\}$ for which $\mathcal{X} \setminus \{A, B, C\}$: $A \sim B$ holds or there exists $D \in \mathcal{X}$ such that $\mathcal{X} \setminus \{A, B, D\}$: $[] \mapsto D$ hold. Then $\{A, B\} \notin \mathcal{C}_s^+(\mathcal{X} \setminus C)$, or $A \notin \mathcal{C}_c^+(\mathcal{X} \setminus B)$ or $B \notin \mathcal{C}_c^+(\mathcal{X} \setminus A)$.

To prove the other direction assume that there exists $C \in \mathcal{X} \setminus \{A, B\}$ such that $\{A, B\} \notin \mathcal{C}_s^+(\mathcal{X} \setminus C)$, or $A \notin \mathcal{C}_c^+(\mathcal{X} \setminus B)$ or $B \notin \mathcal{C}_c^+(\mathcal{X} \setminus A)$. Therefore, $\exists D \in \mathcal{X} \setminus C$, such that $\mathcal{X} \setminus \{A, B, C, D\}$: $A \sim B$ or $\mathcal{X} \setminus \{A, B, C\}$: $[] \mapsto C$ or $\exists D \in \mathcal{X} \setminus B$ such that $\mathcal{X} \setminus \{\mathsf{A}, \mathsf{B}, \mathsf{D}\}$: $[] \mapsto \mathsf{D}$ or $\exists \mathsf{E} \in \mathcal{X} \setminus \mathsf{A}$ such that $\mathcal{X} \setminus \{\mathsf{A}, \mathsf{B}, \mathsf{E}\}$: $[] \mapsto \mathsf{E}$. Hence, by Aug–II, Propagate and Lemma 6, $\mathcal{X} \setminus \{\mathsf{A}, \mathsf{B}\}$: $\mathsf{A} \sim \mathsf{B}$ is not minimal. □

By Lemma 7, the steps in Line 2, 10, 11 and 12 guarantee that the algorithm adds to $\mathcal{M}$ only the minimal ODs of the form $\mathcal{X} \setminus \mathsf{A}$: $[] \mapsto \mathsf{A}$, where $\mathcal{X} \in \mathcal{L}_l$ and $\mathsf{A} \in \mathcal{X}$. By Lemma 8, the steps in Line 4, 6, 17, 18, 20 and 21 ensure that the algorithm adds to $\mathcal{M}$ only the minimal ODs of the form $\mathcal{X} \setminus \{\mathsf{A}, \mathsf{B}\}$: $\mathsf{A} \sim \mathsf{B}$, where $\mathcal{X} \in \mathcal{L}_l$, and $\mathsf{A}, \mathsf{B} \in \mathcal{X}$ and $\mathsf{A} \neq \mathsf{B}$.

The algorithm *computeODs*($\mathcal{L}_l$) also computes the sets $\mathcal{C}_c^+(\mathcal{X})$ and $\mathcal{C}_s^+(\mathcal{X})$ for all $\mathcal{X} \in \mathcal{L}_l$. The following lemma shows that computing $\mathcal{C}_c^+(\mathcal{X})$ is done correctly [11, 10].

LEMMA 9. *Let $\mathcal{C}_c^+(\mathcal{Y})$ be correctly computed $\forall \mathcal{Y} \in \mathcal{L}_{l-1}$. The algorithm computeODs($\mathcal{L}_l$) calculates correctly $\mathcal{C}_c^+(\mathcal{X})$, $\forall \mathcal{X} \in \mathcal{L}_l$.*

We prove in Lemma 10 that computing the candidate set $\mathcal{C}_s^+(\mathcal{X})$ is performed correctly.

LEMMA 10. *Let $\mathcal{C}_c^+(\mathcal{Y})$ and Let $\mathcal{C}_s^+(\mathcal{Y})$ be correctly computed $\forall \mathcal{Y} \in \mathcal{L}_{l-1}$. The algorithm computeODs($\mathcal{L}_l$) computes correctly $\mathcal{C}_s^+(\mathcal{X})$, $\forall \mathcal{X} \in \mathcal{L}_l$.*

**Proof**
A pair of attributes $\{\mathsf{A}, \mathsf{B}\}$, such that $\{\mathsf{A}, \mathsf{B}\} \in \mathcal{X}^2$ and $\mathsf{A} \neq \mathsf{B}$, is in $\mathcal{C}_s^+(\mathcal{X})$ after the execution of *computeODs*($\mathcal{L}_l$) unless it is excluded from $\mathcal{C}_s^+(\mathcal{X})$ on Line 6, 19 or 22. First, we show that if $\{\mathsf{A}, \mathsf{B}\}$ is excluded from $\mathcal{C}_s^+(\mathcal{X})$ by *computeODs*($\mathcal{L}_l$), then $\{\mathsf{A}, \mathsf{B}\} \notin \mathcal{C}_s^+(\mathcal{X})$ by the definition of $\mathcal{C}_s^+(\mathcal{X})$.

– When $\{\mathsf{A}, \mathsf{B}\}$ is excluded from $\mathcal{C}_s^+(\mathcal{X})$ on Line 6, this means that there exists $\mathsf{C} \in \mathcal{X} \setminus \{\mathsf{A}, \mathsf{B}\}$ with $\{\mathsf{A}, \mathsf{B}\} \notin \mathcal{C}_s^+(\mathcal{X} \setminus \mathsf{C})$. Hence, $\mathcal{X} \setminus \{\mathsf{A}, \mathsf{B}, \mathsf{C}\}$: $\mathsf{A} \sim \mathsf{B}$ or there exists $\mathsf{D} \in \mathcal{X} \setminus \mathsf{C}$ such that $\mathcal{X} \setminus \{\mathsf{A}, \mathsf{B}, \mathsf{C}, \mathsf{D}\}$ hold. Hence, $\{\mathsf{A}, \mathsf{B}\} \notin \mathcal{C}_s^+(\mathcal{X})$ by the definition of $\mathcal{C}_s^+(\mathcal{X})$.
– If $\{\mathsf{A}, \mathsf{B}\}$ is excluded on Line 19 this means that $\mathsf{A} \notin \mathcal{C}_c^+(\mathcal{X} \setminus \mathsf{B})$ or $\mathsf{B} \notin \mathcal{C}_c^+(\mathcal{X} \setminus \mathsf{B})$. Hence, there exists $\mathsf{C} \in \mathcal{X} \setminus \mathsf{B}$ such that $\mathcal{X} \setminus \{\mathsf{A}, \mathsf{B}\}$: $[] \mapsto \mathsf{C}$ or there exists $\mathsf{C} \in \mathcal{X} \setminus \mathsf{A}$ such that $\mathcal{X} \setminus \{\mathsf{A}, \mathsf{B}\}$: $[] \mapsto \mathsf{C}$. Hence, $\{\mathsf{A}, \mathsf{B}\} \notin \mathcal{C}_s^+(\mathcal{X})$ by the definition of $\mathcal{C}_s^+(\mathcal{X})$.
– When $\{\mathsf{A}, \mathsf{B}\}$ is excluded on Line 22 then $\mathcal{X} \setminus \{\mathsf{A}, \mathsf{B}\}$ $\mathsf{A} \sim \mathsf{B}$ hold. Hence, $\{\mathsf{A}, \mathsf{B}\} \notin \mathcal{C}_s^+(\mathcal{X})$ by the definition of $\mathcal{C}_s^+(\mathcal{X})$.

Next, we show the other direction, that if $\{\mathsf{A}, \mathsf{B}\} \notin \mathcal{C}_s^+(\mathcal{X})$ by the definition of $\mathcal{C}_s^+(\mathcal{X})$, then $\mathsf{A}$ is excluded from $\mathcal{C}_s^+(\mathcal{X})$ by *computeODs*($\mathcal{L}_l$). Assume $\{\mathsf{A}, \mathsf{B}\} \notin \mathcal{C}_s^+(\mathcal{X})$ by the definition of $\mathcal{C}_s^+(\mathcal{X})$. Therefore, there exists $\mathsf{C} \in \mathcal{X}$ such that $\mathcal{X} \setminus \{\mathsf{A}, \mathsf{B}, \mathsf{C}\}$: $\mathsf{A} \sim \mathsf{B}$ or there exists $\mathsf{D} \in \mathcal{X}$ such that $\mathcal{X} \setminus \{\mathsf{A}, \mathsf{B}, \mathsf{D}\}$: $[] \mapsto \mathsf{D}$.

– When $\mathcal{X} \setminus \{\mathsf{A}, \mathsf{B}, \mathsf{C}\}$: $\mathsf{A} \sim \mathsf{B}$ there are two cases. If $\mathsf{C} = \mathsf{A}$ or $\mathsf{C} = \mathsf{B}$, then $\mathcal{X} \setminus \{\mathsf{A}, \mathsf{B}\}$: $\mathsf{A} \sim \mathsf{B}$ hold, and $\{\mathsf{A}, \mathsf{B}\}$ is removed on Line 22, if $\mathcal{X} \setminus \{\mathsf{A}, \mathsf{B}\}$: $\mathsf{A} \sim \mathsf{B}$ is minimal, and on Line 6 or 19 otherwise. However, if $\mathsf{C} \neq \mathsf{A}, \mathsf{B}$ then $\{\mathsf{A}, \mathsf{B}\} \notin \mathcal{C}_s^+(\mathcal{X} \setminus \mathsf{C})$ and $\{\mathsf{A}, \mathsf{B}\}$ is removed on Line 6.
– If $\mathcal{X} \setminus \{\mathsf{A}, \mathsf{B}, \mathsf{D}\}$: $[] \mapsto \mathsf{D}$ then $\{\mathsf{A}, \mathsf{B}\}$ is excluded on Line 6 or 19.

□

## 4.5 Pruning Levels and Completness

Algorithm 4 shows *pruneLevels*($\mathcal{L}_l$), which implements an additional optimization. We delete node $\mathcal{X}$ from $\mathcal{L}_l$, where $l \geq 2$, if both $\mathcal{C}_c^+(\mathcal{X}) = \{\}$ and $\mathcal{C}_s^+(\mathcal{X}) = \{\}$. Since for all $\mathcal{Y}$ such that $\mathcal{Y} \supset \mathcal{X}$, $\mathcal{C}_c^+(\mathcal{Y}) = \{\}$ and $\mathcal{C}_s^+(\mathcal{Y}) = \{\}$, we can prove the following lemma.

**Algorithm 4** pruneLevels($\mathcal{L}_l$)

```
for all X ∈ L_l do
    if l ≥ 2 then
        if C_c^+(X) = {} and C_s^+(X) = {} then
            Delete X from L_l
        end if
    end if
end for
```

LEMMA 11. *Deleting $\mathcal{X}$ from $\mathcal{L}_l$ for levels $l \geq 2$, if $\mathcal{C}_c^+(\mathcal{X}) = \{\}$ and $\mathcal{C}_s^+(\mathcal{X}) = \{\}$ has no effect on the output set of minimal ODs $\mathcal{M}$.*

**Proof**
If $\mathcal{C}_c^+(\mathcal{X}) = \{\}$ and $\mathcal{C}_s^+(\mathcal{X}) = \{\}$, the loops on Lines 10–14 and Lines 17–22 in the algorithm *computeODs*($\mathcal{L}_l$) will not be executed at all. Consider $\mathcal{Y} \supset \mathcal{X}$ and assume that $\mathsf{A}, \mathsf{B} \in \mathcal{X}$, where $\mathsf{A} \neq \mathsf{B}$, and $\mathsf{C} \in \mathcal{Y} \setminus \mathcal{X}$. The ODs $\mathcal{Y} \setminus \mathsf{C}$: $[] \mapsto \mathsf{C}$ and $\mathcal{Y} \setminus \mathsf{A}$: $[] \mapsto \mathsf{A}$ are not minimal by Augmentation and Lemma 5. Also, the ODs $\mathcal{Y} \setminus \{\mathsf{A}, \mathsf{B}\}$: $\mathsf{A} \sim \mathsf{B}$ and $\mathcal{Y} \setminus \{\mathsf{A}, \mathsf{C}\}$: $\mathsf{A} \sim \mathsf{C}$ (or $\mathcal{Y} \setminus \{\mathsf{A}, \mathsf{C}\}$: $\mathsf{C} \sim \mathsf{A}$) are not minimal by Augmentation-II, Propagate and Lemma 6. Therefore, for all $\mathcal{Y} \supset \mathcal{X}$, $\mathcal{C}_c^+(\mathcal{Y}) = \{\}$ and $\mathcal{C}_s^+(\mathcal{Y}) = \{\}$. Hence, deleting $\mathcal{X}$ from $\mathcal{L}_l$ has no effect on the output set of ODs $\mathcal{M}$ of the algorithm. □

EXAMPLE 11. *Let* $\mathsf{A}$: $[] \mapsto \mathsf{B}$, $\mathsf{B}$: $[] \mapsto \mathsf{A}$ *and* $\{\}$: $\mathsf{A} \sim \mathsf{B}$. *Since $\mathcal{C}_c^+(\{\mathsf{A}, \mathsf{B}\}) = \{\}$ and $\mathcal{C}_s^+(\{\mathsf{A}, \mathsf{B}\})$ as well as $l = 2$, by the pruning levels rule (Lemma 11), the node $\{\mathsf{A}, \mathsf{B}\}$ is deleted and the node $\{\mathsf{A}, \mathsf{B}, \mathsf{C}\}$ is not considered (see Figure 3). This is justified as* $\{\mathsf{AB}\}$: $[] \mapsto \mathsf{C}$ *is not minimal by Lemma 5,* $\{\mathsf{AC}\}$: $[] \mapsto \mathsf{B}$ *is not minimal by Augmentation–I,* $\{\mathsf{BC}\}$: $[] \mapsto \mathsf{A}$ *is not minimal by Augmentation–I,* $\{\mathsf{C}\}$: $\mathsf{A} \sim \mathsf{B}$ *is not minimal by Augmentation–II,* $\{\mathsf{A}\}$: $\mathsf{B} \sim \mathsf{C}$ *is not minimal by Propagate, and* $\{\mathsf{B}\}$: $\mathsf{A} \sim \mathsf{C}$ *is not minimal by Propagate.*

THEOREM 8. *The FASTOD algorithm computes a complete, minimal set of ODs $\mathcal{M}$.*

**Proof**
Steps in Line 2, 10, 11 and 12 by Lemma 7 guarantee that the algorithm adds to $\mathcal{M}$ only the minimal ODs of the form $\mathcal{X} \setminus \mathsf{A}$: $[] \mapsto \mathsf{A}$, where $\mathcal{X} \in \mathcal{L}_l$ and $\mathsf{A} \in \mathcal{X}$. Steps in Line 4, 6, 17, 18, 20 and 21 by Lemma 8 guarantee that the algorithm adds to $\mathcal{M}$ only the minimal ODs of the form $\mathcal{X} \setminus \{\mathsf{A}, \mathsf{B}\}$: $\mathsf{A} \sim \mathsf{B}$, where $\mathcal{X} \in \mathcal{L}_l$, and $\mathsf{A}, \mathsf{B} \in \mathcal{X}$ and $\mathsf{A} \neq \mathsf{B}$.

The algorithm *computeODs*($\mathcal{L}_l$) calculates correctly the sets $\mathcal{C}_c^+(\mathcal{X})$ and $\mathcal{C}_s^+(\mathcal{X})$ for all $\mathcal{X} \in \mathcal{L}_l$ by induction given Lemma 9 and 10. Also, by Lemma 11 deleting $\mathcal{X}$ from $\mathcal{L}_l$ for levels $l \geq 2$, if $\mathcal{C}_c^+(\mathcal{X}) = \{\}$ and $\mathcal{C}_s^+(\mathcal{X}) = \{\}$ has no effect on the output set of minimal ODs $\mathcal{M}$ of the algorithm. Hence, the *FASTOD* algorithm computes a sound and complete set of minimal ODs $\mathcal{M}$. □

While we provide theoretical guarantees for *FASTOD* to find a complete set of ODs (Sections 4.4–4.5), this is not the case for the *ORDER* algorithm in [13]. In fact, *ORDER* is sound, but *not* complete as it misses important ODs. When *ORDER* accesses the node $[\mathsf{A}, \mathsf{B}, \mathsf{C}]$ in the list-containment lattice, it generates the list-based ODs $[\mathsf{B}, \mathsf{C}] \mapsto [\mathsf{A}]$ and $[\mathsf{C}] \mapsto [\mathsf{A}, \mathsf{B}]$. However, it discards potential ODs with repeating attributes in a dependency. For instance, a valid

OD $[\mathsf{C}] \mapsto [\mathsf{C}, \mathsf{A}, \mathsf{B}]$ (note, an FD) is missed, if $[\mathsf{C}] \sim [\mathsf{A}, \mathsf{B}]$ does not hold. Similarly, a valid OD $[\mathsf{C}] \sim [\mathsf{A}, \mathsf{B}]$ (an order compatible dependency) is missed, if $[\mathsf{C}] \mapsto [\mathsf{C}, \mathsf{A}, \mathsf{B}]$ does not hold. This causes the algorithm to discard ODs such as d_month $\sim$ d_week from Example 2 as d_month does not *functionally determine* d_week. Only if both $[\mathsf{C}] \mapsto [\mathsf{C}, \mathsf{A}, \mathsf{B}]$ and $[\mathsf{C}] \sim [\mathsf{A}, \mathsf{B}]$ hold, then they are not missed, as just $[\mathsf{C}] \mapsto [\mathsf{A}, \mathsf{B}]$ is checked by the *ORDER* algorithm. (By Theorem 1, $[\mathsf{C}] \mapsto [\mathsf{A}, \mathsf{B}]$ *iff* $[\mathsf{C}] \mapsto [\mathsf{C}, \mathsf{A}, \mathsf{B}]$ and $[\mathsf{C}] \sim [\mathsf{A}, \mathsf{B}]$.) Also, *ORDER* misses potential ODs with the same prefix; i.e., $\mathbf{XY} \mapsto \mathbf{XZ}$, such as [year, salary] $\mapsto$ [year, bin] (Example 4). Furthermore, *ORDER* discards constants; i.e., ODs in the form of $[\,] \mapsto \mathbf{X}$. For instance, if all flight data are from the year 2012, the year attribute is a constant (see Sections 5). This inadvertent incompleteness, allows the authors in [13] to devise aggressive pruning rules to overcome partially the factorial cost. However, this comes at a price of incompleteness, of discarding many important ODs.

## 4.6 Efficient OD Validation

**Computing with Partitions.** In order to verify whether the ODs $\mathcal{X}$: $[\,] \mapsto \mathsf{A}$ and $\mathcal{X}$: $\mathsf{A} \sim \mathsf{B}$ hold, we first compute the equivalence classes over the context $\mathcal{X}$, i.e., $\Pi_{\mathcal{X}}$. Next, within each equivalence class in $\Pi_{\mathcal{X}}$, we verify whether the ODs $\mathcal{X}$: $[\,] \mapsto \mathsf{A}$ and $\mathcal{X}$: $\mathsf{A} \sim \mathsf{B}$ hold, respectively.

The values of the columns are replaced with integers: 1, 2, ..., $n$, in a way that the equivalence classes do not change and the ordering is preserved. That is, the same values are substituted by the same integers, and higher values are replaced by larger integers. Computation over integers is more time and space efficient. After translating values to integers, the value $t_{\mathsf{A}}$ is used as the identifier of the equivalence class $\mathcal{E}(t_{\mathsf{A}})$ of $\Pi_{\mathsf{A}}$.

After computing the partitions $\Pi_{\mathsf{A}}$, $\Pi_{\mathsf{B}}$, ..., for all single attributes in $\mathbf{R}$ at level $l = 2$, we efficiently compute the partitions for subsequent levels in linear time by taking the product of refined partitions, i.e., $\Pi_{\mathsf{A} \bigcup \mathsf{B}} = \Pi_{\mathsf{A}} \cdot \Pi_{\mathsf{B}}$. Accordingly, for all other levels the partitions are not calculated from scratch for each set of attributes $\mathcal{X}$. In the general case, for all $|\mathcal{X}| \geq 2$, partitions $\Pi_{\mathcal{X}}$ are computed in linear time as products of partitions $\Pi_{\mathcal{Y}}$ and $\Pi_{\mathcal{Z}}$, i.e., $\Pi_{\mathcal{X}} = \Pi_{\mathcal{Y}} \cdot \Pi_{\mathcal{Z}}$, such that $\mathcal{Y}, \mathcal{Z} \subset \mathcal{X}$ and $|\mathcal{Y}| = |\mathcal{Z}| = |\mathcal{X} - 1|$. This is beneficial for a levelwise algorithm since only partitions from the previous level are needed. Hence, this optimizes both time and space.

To verify whether an OD $\mathcal{X}$: $[\,] \mapsto \mathsf{A}$ holds, for each equivalence class $\mathcal{E}(t_{\mathcal{X}}) \in \Pi_{\mathcal{X}}$, we need to check whether $|\Pi_{\mathsf{A}}(\mathcal{E}(t_{\mathcal{X}}))| = 1$. Therefore, this step can be accomplished in linear time in the number of tuples and can be optimized by a technique called *error rate* used in [11] to avoid potentially scans over the dataset.

Verifying if an OD $\mathcal{X}$: $\mathsf{A} \sim \mathsf{B}$ holds is more involved but can also be computed efficiently. For all single attributes $\mathsf{A} \in \mathbf{R}$ at level $l = 2$, we calculate *sorted partitions* $\tau_{\mathsf{A}}$. A sorted partition $\tau_{\mathsf{A}}$ is a list of equivalence classes according to the ordering imposed on the tuples by A. For instance, in Table 1, $\tau_{\mathsf{bin}} = \{\{t1, t4\}, \{t2, t5\}, \{t3, t6\}\}$. Hence, when we verify for any subsequent level, whether $\mathcal{X}$: $\mathsf{A} \sim \mathsf{B}$ is valid, for each equivalence class $\mathcal{E}(t_{\mathcal{X}}) \in \Pi_{\mathcal{X}}$, we compute $\tau_{\mathsf{A}}(\mathcal{E}(t_{\mathcal{X}}))$ by hashing tuples into sorted buckets with a single scan over $\tau_{\mathsf{A}}$. (See the example in Table 2.) This allows us to verify efficiently whether there is no swap over attributes A and B for each equivalence class in $\Pi_{\mathcal{X}}$ via a single scan

| $\tau_{\mathsf{A}}$ Rank | Tuple #'s |
|---|---|
| 1 | $\{t3, t5, t8\}$ |
| 2 | $\{t1, t6\}$ |
| 3 | $\{t4\}$ |
| 4 | $\{t7\}$ |
| 5 | $\{t2\}$ |

(a) $\tau_{\mathsf{A}}$

| $\Pi_{\mathcal{X}}$ ID | $\tau_{\mathsf{A}}(\mathcal{E}(t_{\mathcal{X}}))$ |
|---|---|
| $t1_{\mathcal{X}}$ | $\{t1\}$ |
| $t2_{\mathcal{X}}$ | $\{t2\}$ |
| $t3_{\mathcal{X}}$ | $\{t3, t5\}, \{t4\}$ |
| $t6_{\mathcal{X}}$ | $\{t6\}, \{t7\}$ |
| $t8_{\mathcal{X}}$ | $\{t8\}$ |

(b) $\tau_{\mathsf{A}}(\mathcal{E}(t_{\mathcal{X}}))$ for each equivalence class

Table 2: $\Pi_{\mathcal{X}} = \{\{\texttt{t1}\}, \{\texttt{t2}\}, \{\texttt{t3}, \texttt{t4}, \texttt{t5}\}, \{\texttt{t6}, \texttt{t7}\}, \{\texttt{t8}\}\}$, $\tau_A = \{\{\texttt{t3}, \texttt{t5}, \texttt{t8}\}, \{\texttt{t1}, \texttt{t6}\}, \{\texttt{t4}\}, \{\texttt{t7}\}, \{\texttt{t2}\}\}$

over the dataset. Hence, checking whether $\mathcal{X}$: $\mathsf{A} \sim \mathsf{B}$ can be done in linear time.

**Key Pruning.** When a key is found during the search of ODs, additional optimization methods can be applied. Call a set of attributes $\mathcal{X}$ a *superkey* if no two tuples in the relation instance $\mathbf{r}$ agree on $\mathcal{X}$. $\mathcal{X}$ is a *key* if it is a superkey and no proper subset of it is a superkey.

Lemma 12. *Let $\mathcal{X} \setminus \mathsf{A}$ be a superkey and $\mathsf{A} \in \mathcal{X}$. The OD $\mathcal{X} \setminus \mathsf{A}$: $[\,] \mapsto \mathsf{A}$ is valid. The OD $\mathcal{X} \setminus \mathsf{A}$: $[\,] \mapsto \mathsf{A}$ is minimal iff $\mathcal{X} \setminus \mathsf{A}$ is a key and $\forall \mathsf{B} \in \mathcal{X}$, $\mathsf{A} \in \mathcal{C}_c^+(\{\mathcal{X}\} \setminus \mathsf{B})$.*

**Proof**
Since $\mathcal{X} \setminus \mathsf{A}$ is a superkey, partition $\Pi_{\mathcal{X} \setminus \mathsf{A}}$ consists of singleton equivalence classes only. Hence, the OD $\mathcal{X} \setminus \mathsf{A}$: $[\,] \mapsto \mathsf{A}$ is valid. The remaining follows from Lemma 7. □

Typically, an OD $\mathcal{X} \setminus \mathsf{A}$: $[\,] \mapsto \mathsf{A}$, $\mathsf{A} \notin \mathcal{X}$, is verified on Line 11 in the algorithm *computeODs*$(\mathcal{L}_l)$. However, if $\mathcal{X} \setminus \mathsf{A}$ is a key, then $\mathcal{X} \setminus \mathsf{A}$: $[\,] \mapsto \mathsf{A}$ always holds, and hence, we do not need to verify it. On the other hand, if $\mathcal{X} \setminus \mathsf{A}$ is a superkey but not a key, then clearly the OD $\mathcal{X} \setminus \mathsf{A}$: $[\,] \mapsto \mathsf{A}$ is not minimal. This is because there exists $\mathsf{B} \in \mathcal{X}$, such that $\mathcal{X} \setminus \mathsf{B}$ is a superkey, therefore, $\mathsf{A} \notin \mathcal{C}_c^+(\{\mathcal{X}\} \setminus \mathsf{B})$).

Similarly, Lemma 13 states that $\mathcal{X} \setminus \{\mathsf{A}, \mathsf{B}\}$: $\mathsf{A} \sim \mathsf{B}$, $\mathsf{A}$, $\mathsf{B} \in \mathcal{X}$, also does not have to be verified in Line 20 of the algorithm *computeODs*$(\mathcal{L}_l)$, if $\mathcal{X} \setminus \{\mathsf{A}, \mathsf{B}\}$ is a key since it is not minimal (this follows from the Propagate axiom).

Lemma 13. *Let $\mathcal{X}$ be a key (or superkey) and $\mathsf{A}, \mathsf{B} \in \mathbf{R} \setminus \mathcal{X}$. The OD $\mathcal{X}$: $\mathsf{A} \sim \mathsf{B}$ is valid but not minimal.*

**Proof**
As $\mathcal{X}$ is a key (or superkey), partition $\Pi_{\mathcal{X}}$ consists of singleton equivalence classes only. Therefore, the OD $\mathcal{X}$: $\mathsf{A} \sim \mathsf{B}$ holds. By Lemma 12, the OD $\mathcal{X}$: $[\,] \mapsto \mathsf{A}$ holds since $\mathcal{X}$ is a key (or superkey). Hence, $\{\mathsf{A}, \mathsf{B}\} \notin \mathcal{C}_s^+(\mathcal{X} \bigcup \{\mathsf{A}, \mathsf{B}\})$. □

**Stripped Partitions.** We replace partitions, with a more compact version, called *stripped partitions.* A stripped partition is a partition with equivalence classes of size one–call these singleton equivalence classes–discarded. A stripped representation of partition $\Pi_{\mathcal{X}}$ is denoted as $\Pi^*_{\mathcal{X}}$.

Example 12. *In Table 1,* $\Pi^*_{\mathsf{salary}} = \{\{t2, t6\}\}$*, whereas* $\Pi_{\mathsf{salary}} = \{\{t1\}, \{t2, t6\}, \{t3\}, \{t4\}, \{t5\}\}$.

Removing singleton equivalence classes is correct as they cannot break any set-based OD in the form of $\mathcal{X}$: $[\,] \mapsto \mathsf{A}$ or $\mathcal{X}$: $\mathsf{A} \sim \mathsf{B}$.

Lemma 14. *Singleton equivalence classes over attribute set $\mathcal{X}$ cannot falsify any OD: $\mathcal{X}$: $[\,] \mapsto \mathsf{A}$ or $\mathcal{X}$: $\mathsf{A} \sim \mathsf{B}$.*

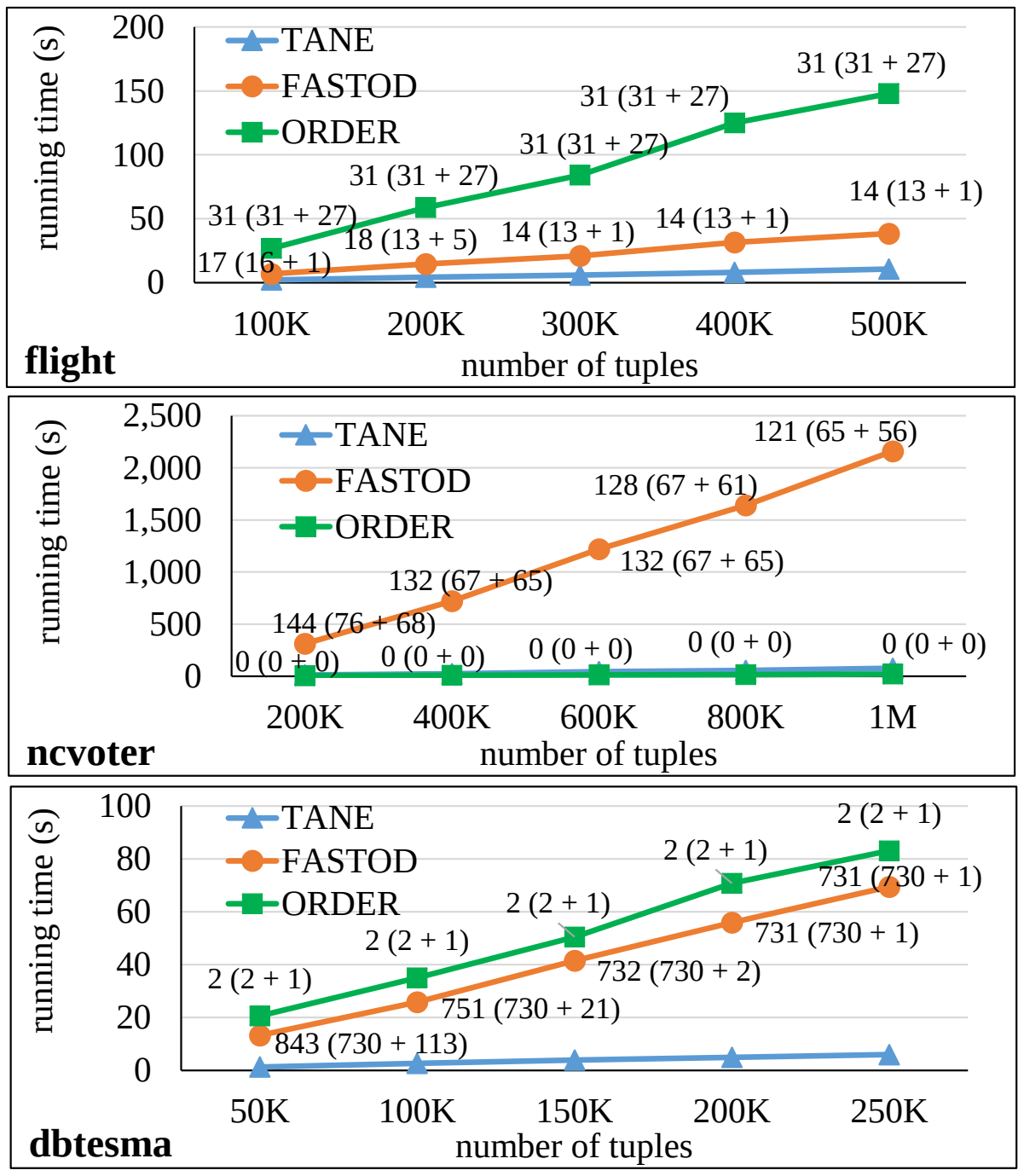

Figure 4: Scalability and effectiveness in $|\mathbf{r}|$.

**Proof**

Follows directly from the definition of context (Def. 6). □

If $\Pi^*_{\mathcal{X}} = \{\}$, then $\mathcal{X}$ is a key (or superkey) and the optimization with key pruning described above is triggered.

Note that, unlike in *FASTOD*, stripped partitions are not used in *ORDER* [13]. The authors in [13] explicitly state that stripped partitions cannot be used with their list-based canonical form.

## 4.7 Complexity Analysis

The complexity of most data dependency discovery algorithms depends on the number of candidates in the lattice. The previous state-of-the-art discovery algorithm, *ORDER*, for ODs has a *factorial* worst-case complexity in the number of attributes [13], as it traverses a lattice of attribute *permutations* with $\lfloor |\mathbf{R}|! * e \rfloor$ nodes. The authors in [13] claim that this is inevitable, due to the factorial search space of candidates, since ODs are defined over lists of attributes as opposed to FDs being defined over sets of attributes.

However, we show that the worst-case complexity of our OD discovery algorithm is *exponential* in the number of attributes, as there exists a quadratic mapping of list-based ODs into the equivalent set-based ODs (Section 3.1), and there are $2^{|\mathbf{R}|}$ nodes in the set-containment lattice (Sections 4.1 and 4.3). This establishes a lower bound as ODs subsume FDs, and it has been already observed for FDs that the solution space is exponential, and a polynomial time algorithm in the number of attributes cannot exist [11]. Additionally, the complexity is linear in the number of tuples (Sections 4.6). The linearity makes *FASTOD* especially suitable for relations with a large number of tuples. These results are in line with the previous FD and inclusion dependency algorithms [11].

# 5. EXPERIMENTS

In this section, we present an experimental evaluation of our techniques, with a focus on the following objectives.

1. An *evaluation* of the efficiency of our approach using real and synthetic datasets. This includes scalability and effectiveness over different problem characteristics: size of the dataset and number of attributes (Section 5.2) as well as the level of the lattice (Section 5.4).
2. A *comparison* against *ORDER*, the OD discovery algorithm from [13], quantifying the benefits of our approach in terms of time, conciseness and completeness (Section 5.3). We also compare our algorithm with the FD discovery algorithm [11].
3. An *evaluation* of the effectiveness of our pruning optimizations from Section 4.2 and Section 4.5 (Section 5.4).

Our experiments were run on a machine with an Intel Xeon CPU E5-2630 v3 2.4GHz with 64GB of memory. The algorithm was implemented in Java 8.

## 5.1 Data Characteristics

We use several real and synthetic datasets that have previously been used to evaluate FD and OD discovery algorithms [13]. They are published through the UCI Machine Learning Repository[3] and the Hasso-Plattner-Institute (HPI) repository[4]. The `flight` dataset (from the HPI repository) contains information about US domestic flights, with 500K tuples and 40 attributes. The `ncvoter` dataset (from the UCI repository) contains personal data of registered voters from North Carolina, with 1M tuples and 20 attributes. The `hepatitis` datasets provides information about the hepatitis disease of the liver, with 155 tuples and 20 attributes. `Dbtesma` is a synthetic dataset with 250K tuples and 30 attributes, available at the link provided for the HPI repository, produced by a data generator[5] for benchmarks and performance analysis.

## 5.2 Scalability

We evaluate the running time of our OD discovery algorithm by varying the number of tuples $|\mathbf{r}|$ and the number of attributes $|\mathbf{R}|$.

**Exp-1: Scalability in $|\mathbf{r}|$.** We measure the running time of *FASTOD* in milliseconds by varying the number of tuples, as reported in Figure 4. Ignore the other curves labeled ORDER and TANE, as well as the numbers next to the datapoints, initially. We use random samples of 20, 40, 60, 80 and 100 percent of `flight`, `ncvoter` and `dbtesma`, all with 10 attributes. The reported runtimes are averaged over ten executions.

Figure 4 shows a linear runtime growth as the computation is dominated by the verification of set-based ODs which requires a scan of the dataset. Therefore, our algorithm scales well for large datasets. Furthermore, it appears to run faster on the `flight` dataset than on the `ncvoter` dataset. This is due to the varying effectiveness of our pruning strategies (details in Section 5.4).

**Exp-2: Scalability in $|\mathbf{R}|$.** We measure the running time of *FASTOD* in milliseconds by varying the number of attributes by taking random projections of the tested datasets. We use the **flight** dataset with 1K tuples, the **ncvoter** dataset with 1K tuples, the **hepatitis** dataset with 155 tuples and the **dbtesma** dataset with 1K tuples. Again, we report the average runtimes over ten executions. Figure 5 show that the running time increases exponentially with the number of attributes. The Y axis of Figure 5 is in log scale. This is not surprising because the number of minimal ODs over the set-containment lattice is exponential in the worst case (Section 4.7).

[3] http://archive.ics.uci.edu/ml
[4] http://metanome.de
[5] http://sourceforge.net/projects/dbtesma

## 5.3 Comparative Study

**Exp-3: Comparison with ORDER [13].** We now compare *FASTOD* with *ORDER* from [13]. We obtained a Java implementation of *ORDER* from www.metanome.de. As Figures 4 and 5 show, our algorithm can be orders of magnitude faster. For instance, on the **flight** dataset with 1K tuples and 20 attributes, *FASTOD* finishes the computation in less than 1 second, whereas *ORDER* did not terminate after 5 hours. (We aborted the experiments after 5 hours and denote it in figures as "* 5h".) The running time of *ORDER* is consistent with results reported in [13]. Similar observations can be made on the **dbtesma** dataset. This is expected as the worst-time complexity for *ORDER* is factorial in the number of attributes [13], whereas, for *FASTOD*, it is exponential in the number of attributes (Section 4.7).

The higher complexity of *ORDER* is remedied in some cases by its aggressive pruning strategies. For instance, the *swap pruning rule* states that if $\mathbf{X} \mapsto \mathbf{Y}$ is invalidated by a swap, then $\mathbf{X}\mathsf{A} \mapsto \mathbf{Y}\mathsf{B}$ is not valid and is not considered when traversing the list-containment lattice. As a result, the *ORDER* algorithm is *not* complete (recall Section 4.5). For example, it misses ODs in the form of $\mathbf{X}\mathsf{A} \mapsto \mathbf{X}\mathsf{A}\mathbf{Y}\mathsf{B}$, e.g., $\mathcal{X}\mathsf{A}$: $[] \mapsto \mathsf{B}$ is missed. It also skips ODs in the form of $\mathbf{X} \mapsto \mathbf{X}\mathbf{Y}$. Similarly, if $\mathbf{X}\mathsf{A} \mapsto \mathbf{Y}\mathsf{B}$ is invalidated by a split, then $\mathbf{X}\mathsf{A} \sim \mathbf{Y}\mathsf{B}$ is not considered; e.g., $\mathcal{X}\mathcal{Y}$: $\mathsf{A} \sim \mathsf{B}$ is missed.

These observations explain why *ORDER* is faster than *FASTOD* (and even *TANE*) on the **hepatitis** dataset: here, *ORDER* discovers zero ODs, whereas *FASTOD*, which is sound and complete, discovers hundreds. Similar reasoning explains the runtimes over the **ncvoter** dataset.

While the swap pruning rule (and other pruning rules) in the *ORDER* algorithm can be deactivated to enable completeness, this has an extreme performance impact. For example, after disabling the swap pruning rule, the *ORDER* algorithm did not terminate within five hours in any of tested datasets. In fact, in [13] they have shown the same.

Furthermore, our canonical representation is significantly more concise than the one in [13] even though *FASTOD* is complete. For brevity, we refer to set-based ODs in the form of $\mathcal{X} \setminus \mathsf{A}$: $[] \mapsto \mathsf{A}$ as FDs and $\mathcal{X} \setminus \{\mathsf{A}, \mathsf{B}\}$: $\mathsf{A} \sim \mathsf{B}$ as order compatible dependencies (OCDs). We report the number of set-based ODs (number of FDs and number of OCDs) in Figure 4 and 5 next to the runtime datapoints. For instance, for the **flight** dataset with 500K tuples and 10 attributes, the number of discovered set-based ODs by *FASTOD* is 14 (13 FDs and 1 OCD) and list-based ODs by *ORDER* is 31, which maps to 58 set-based ODs (31 FDs and 27 OCDs). Since all flight data are from the year 2012, the year attribute is a constant. Therefore, *FASTOD* discovers an OD $\{\}$: $[] \mapsto$ [year]; however, *ORDER* produces ODs [quarter] $\mapsto$ [year], [dayOfMonth] $\mapsto$ [year], [dayOfWeek] $\mapsto$ [year] and so on, which follow from $\{\}$: $[] \mapsto$ [year] by Augmentation–I and

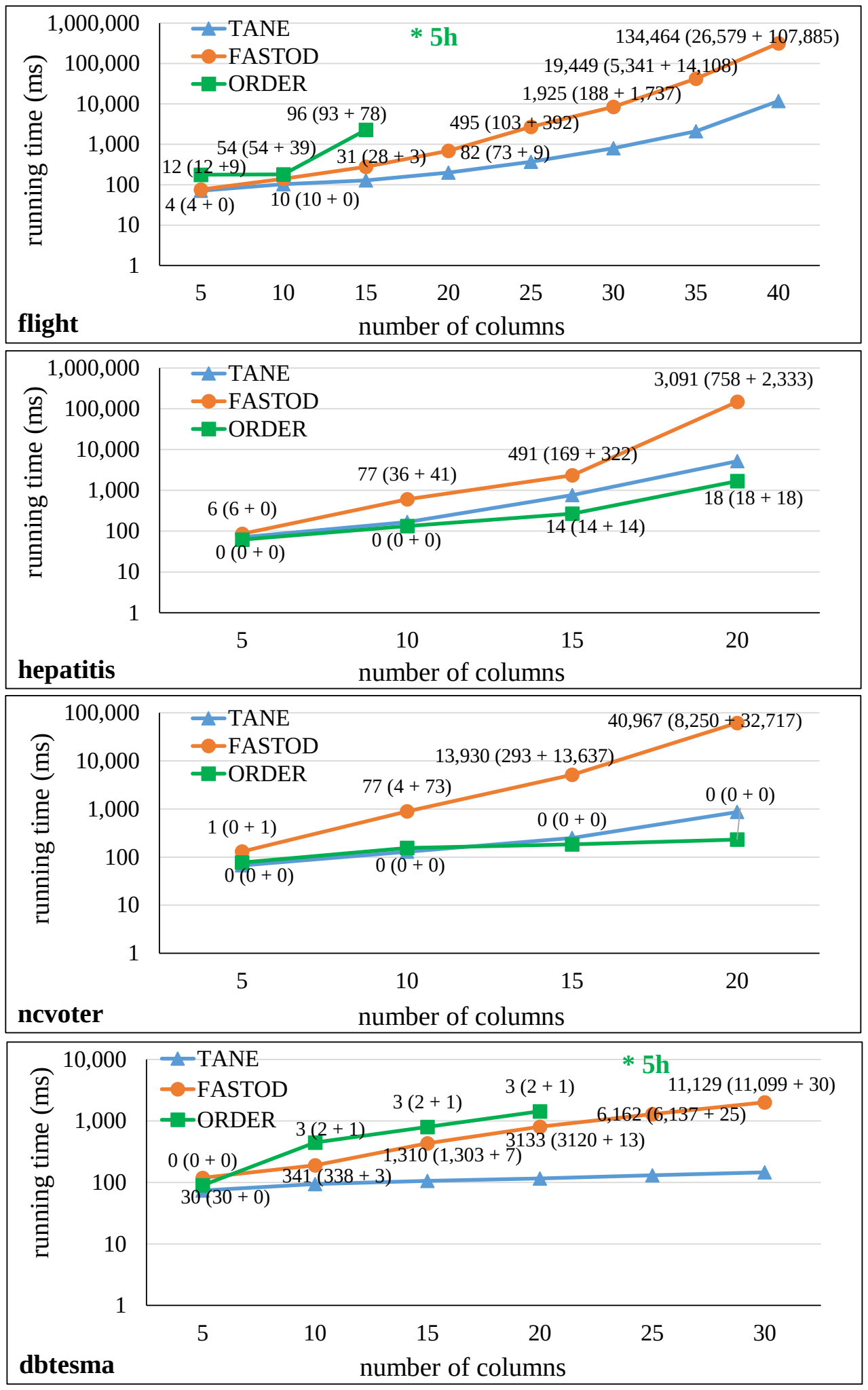

Figure 5: Scalability and effectiveness in $|\mathbf{R}|$.

Propagate (but not vice versa!). For example, $\{$quarter$\}$: $[] \mapsto$ [year] can be inferred from Augmentation–I, and $\{\}$: [quarter] $\sim$ [year] can be deduced from Propagate. Given a query with `order by year`, the ODs found by *FASTOD* can eliminate the sort operator (specifically, the OD stating that year is a constant). On the other hand, since $\{\}$: $[] \mapsto$ [year] does not follow from the ODs discovered by the *ORDER* algorithm, sorting cannot be avoided.

**Exp-4: Comparison with TANE [11].** Since ODs are closely related to FDs, as ODs properly subsume FDs, we also conduct a comparative study with *TANE* [11] (Java implementation). Here, we effectively measure the extra cost to capture the additional OD semantics: those of order. As expected, *TANE* is faster than *FASTOD* (see Figures 4 and 5). TANE uses an effective technique called *error rate* [11] to speed up the verification over FDs (detecting splits) that is not applicable to ODs in the form of $\mathcal{X} \setminus \{\mathsf{A}, \mathsf{B}\}$: $\mathsf{A} \sim \mathsf{B}$ (detecting swaps). However, both *TANE* and *FASTOD* scale linearly in the number tuples and exponentially in the number of attributes. We argue that the extra cost of OD discovery is worthwhile, as ODs can convey many business rules that FDs cannot express. For instance, in the *flight* dataset with 1K tuples and 25 attributes, we found around 500 set-based minimal ODs, out of which around 100 are in the form of $\mathcal{X} \setminus \mathsf{A}$: $[] \mapsto \mathsf{A}$ (FDs) (the number of FDs detected by *TANE* and *FASTOD* is the same) and around 400 are in the form of $\mathcal{X} \setminus \{\mathsf{A}, \mathsf{B}\}$: $\mathsf{A} \sim \mathsf{B}$ (OCDs).

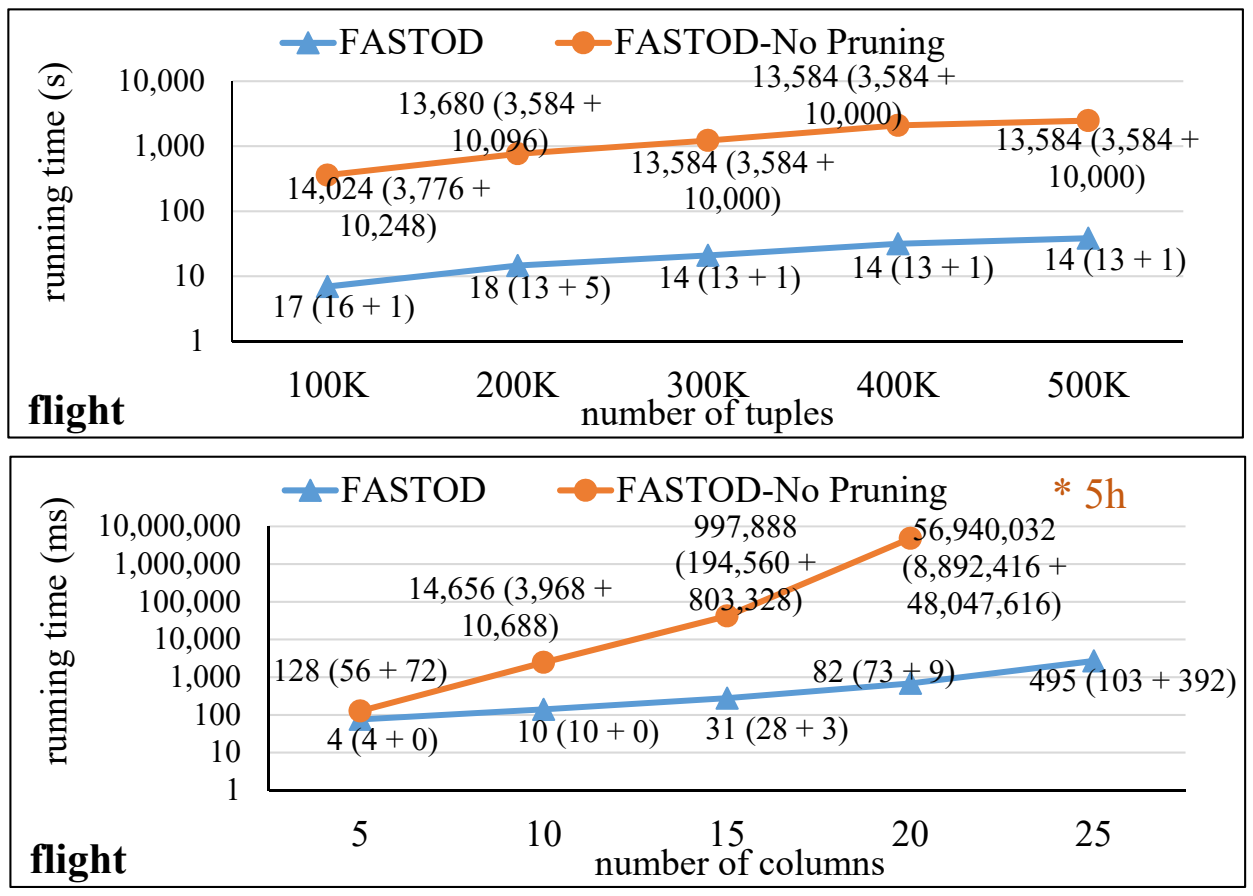

Figure 6: Impact of pruning.

## 5.4 Optimizations and Lattice Levels

**Exp-5: Improving performance**. We now quantify the runtime improvements due to the optimizations described in Sections 4.2 and 4.5. We perform this experiment over the `flight` dataset up to 500K tuples (and 10 attributes) as well as up to 25 attributes (and 1K tuples). Results are shown in Figure 6 for *FASTOD* and *FASTOD-No Pruning*; the Y axis is in log scale. Our optimizations provide a substantial performance improvement. For instance, the running time over the `flight` dataset with 1K tuples and 20 attributes with pruning strategies enabled is less than 1 second versus with disabled pruning strategies, it is around 80 minutes. For the same dataset with 25 attributes the algorithm with optimizations finishes in less than 3 seconds, whereas with pruning strategies deactivated, it does not terminate after 5 hours.

**Exp-6: Pruning non-minimal ODs**. Next, we show that our optimizations prune a significant number of non-minimal ODs. The number of ODs returned by an OD discovery algorithm can be large, but many of them are redundant (non-minimal), as they can be inferred from others. Figure 6 reports the numbers of ODs with and without redundant ODs for the `flight` dataset. Our OD canonical representation is able to prune a large amount of inferred ODs. For instance, over the flight dataset with 1K tuples and 20 attributes, there are around 700 minimal ODs discovered (with pruning activated) and around 50 million non-minimal ODs (with pruning deactivated). Therefore, our canonical representation is highly effective in avoiding redundancy.

**Exp-7: Effectiveness over lattice levels**. Finally, we measure the running time and the number of ODs discovered at different levels of the lattice (Figure 7 with log scale on Y axis). Note that the level number ($l$) determines the size of the context for set-based ODs; i.e., the higher the level, the larger the context. We report results over the `flight` dataset with 1K tuples and 40 attributes. We believe that ODs with a smaller context are more interesting generally as, for instance, they are more likely to be used in query op-timization [24, 25]. Note that FASTOD discards redundant ODs if they can be inferred by ODs detected at lower levels of the lattice. Interestingly, most of the ODs are discovered efficiently over the first few levels of the lattice. Level 9 was the highest level for which FASTOD generated candidates. Since the set-lattice has a shape of a diamond (Figure 3) and nodes are pruned over time, the time to process each level first increases, up to level 6, and decreases thereafter.

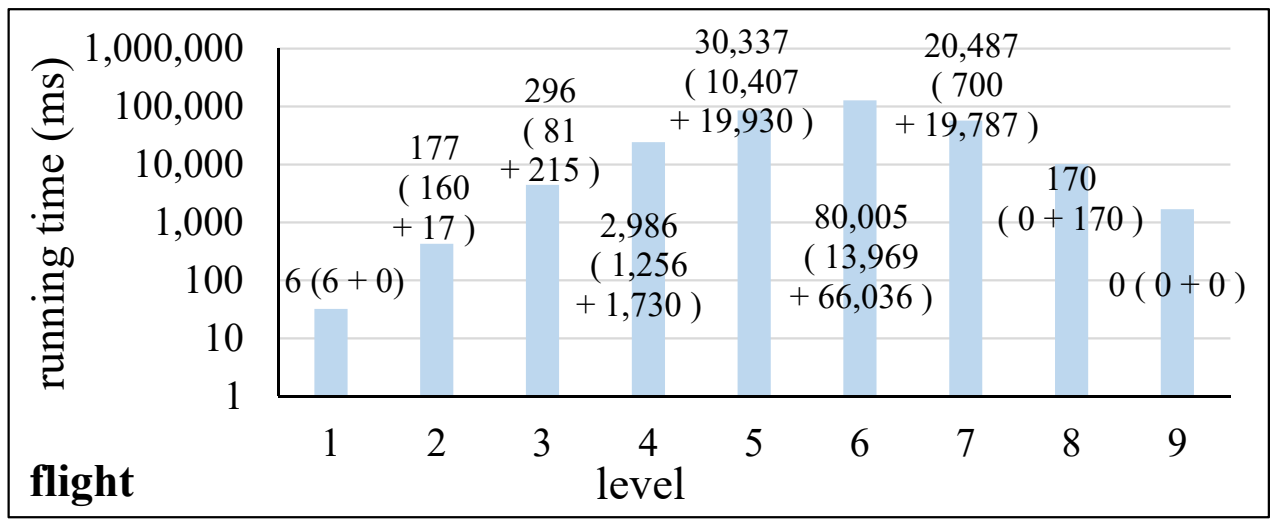

Figure 7: Experiment with lattice levels, $|\mathbf{r}| = 1\text{K}$, $|\mathbf{R}| = 40$.

# 6. RELATED WORK

Pointwise order dependencies (recall Section 2.1) were introduced for the first time in the context of database systems by Ginsburg and Hull [7]. A sound and complete set of axioms was presented, and the inference problem was proven to be co-NP complete. Lexicographical ODs were studied in [17, 22, 25]. Ng [17] developed a theory behind both lexicographical ODs as well as a simpler version of pointwise ODs. A chase procedure was defined for the lexicographical ODs; however, the axiomatization and the complexity of the inference problem for ODs had not been studied. Szlichta et al. [22] presented a sound and complete axiomatization for ODs. Bidirectional ODs, containing a mix of ascending and descending orders, were introduced by Szlichta et al. [25], where it was shown that the inference problem for both ODs and bidirectional ODs is co-NP-complete.

An interesting study to establish whether a given stream is sufficiently nearly-sorted was described by Ben-Moshe et al. [3]. Golab et al. [8] introduced *sequential dependencies* (SDs), which specify that when tuples have consecutive antecedent values, their consequents must be with a specified range. The discovery problem has been studied for approximate SDs [8].

Denial Constraints (DCs) [4], a universally quantified first-order logic formalism, is more expressive than FDs and ODs. However, while the space to be explored for discovery of FDs and ODs is $2^{|\mathbf{R}|}$, the search space for DCs discovery is of size $2^{12*|\mathbf{R}|*(2*|\mathbf{R}|-1)}$, $O(2^{|\mathbf{R}|^2})$. Therefore, the DCs have a much larger space to explore than FDs and ODs. Note that $2^{|\mathbf{R}|^2}$ grows faster than even factorial $|\mathbf{R}|!$ in [13]. The authors in [4] design a set of sound but not complete axioms to develop pruning rules for DC discovery.

There has been much research on FD discovery [11, 10, 18]. However, there has, to date, been little work on discovery of ODs. Langer and Naumann [13] propose the algorithm *ORDER* that uses list-containment lattice with factorial worst-case time complexity in the number of attributes, which establishes an upper bound. However, OD discovery is no harder than FD discovery asymptotically as we show in Section 4.7.

Sorting is at the heart of many database operations; such as sort-merge join, index generation, duplicate elimination and ordering the output for the SQL order-by operator. The importance of sorted sets for query optimization and processing had been long recognized. The seminal query optimizer System R [20] paid particular attention to *interesting orders* by keeping track of all such ordered sets throughout the process of query optimization. FDs and ODs were shown to help generate interesting orders in [21] and in [25, 24], respectively. A practical application of dependencies for improved index design was presented in [6]. In [9], the authors explored the use of sorted sets for executing nested queries. The significance of sorted sets has prompted researchers to look *beyond* the sets that have been explicitly generated. Malkemus et al. [16] and Szlichta et al. [24] showed how to use sorted sets created as generated columns (SQL functions and algebraic expressions) in predicates for business query optimization. Relationships between sorted attributes discovered by reasoning over the physical schema have been also used to eliminate joins over data warehouse queries [23].

# 7. CONCLUSIONS

We have presented an efficient algorithm for discovering order dependencies (ODs) from data, which we showed to be substantially faster than the state-of-the-art. The technical innovation that made our algorithm possible is a novel mapping into a set-based canonical form. We also developed a sound and complete axiomatization for set-based canonical ODs which led to further optimizations to prune the search space of candidate ODs. There is naturally more to be done. In future work, we plan to extend our OD discovery framework to bidirectional ODs [25]. We will also consider the notion of *approximate* ODs that almost hold over a relation instance within a specified threshold. This will enable us to experiment with datasets of varying error rates. Finally, we also plan to study *conditional* ODs that hold over portions of a relation. Since conditional ODs allow data bindings, a large number of individual dependencies may hold on a table, complicating detection of constraint violations.